\documentclass[amsmath,amssymb,prd, notitlepage,nofootinbib,twocolumn, superscriptaddress]{revtex4-2}

\usepackage{bm}

\newcommand{\aap}{{Astron.~Astrophys.}}

\newcommand{\apjl}{{Astrophys.~J.~Lett.}}
\newcommand{\apjs}{{Astrophys.~J.~Supp.}}

\newcommand{\LCDM}{$\Lambda\rm{CDM}$}
\newcommand{\Planck}{{\it{Planck}}}

\newcommand{\Amod}{A_{\rm mod}}

\usepackage{graphicx}
\usepackage[large]{subfigure}
\usepackage{amssymb, amsmath}
\usepackage[amssymb]{SIunits}
\usepackage{aas_macros}
\usepackage{natbib}

\usepackage{xcolor}
\usepackage{booktabs}
\usepackage{multirow} 
\usepackage{verbatim}

\begin{document}

\preprint{APS/123-QED}

\title{Reconstructing the shape of the non-linear matter power spectrum using CMB lensing and cosmic shear}

\author{Karen Perez Sarmiento}\email{kaper@sas.upenn.edu}
\affiliation{Department of Physics and Astronomy, University of Pennsylvania, 209 South 33rd Street, Philadelphia, PA, USA 19104}

\author{Alex Lagu\"e}%
\affiliation{Department of Physics and Astronomy, University of Pennsylvania, 209 South 33rd Street, Philadelphia, PA, USA 19104}%

\author{Mathew S. Madhavacheril}
\affiliation{Department of Physics and Astronomy, University of Pennsylvania, 209 South 33rd Street, Philadelphia, PA, USA 19104}%

\author{Bhuvnesh Jain}
\affiliation{Department of Physics and Astronomy, University of Pennsylvania, 209 South 33rd Street, Philadelphia, PA, USA 19104}%

\author{Blake Sherwin}
\affiliation{Department of Applied Mathematics and Theoretical Astrophysics, University of Cambridge, Cambridge CB3 0WA, United Kingdom}%

\date{\today}
\smallskip

\begin{abstract}
We reconstruct the non-linear matter power spectrum $P(k)$  using a joint analysis of gravitational lensing of the cosmic microwave background (CMB) and lensing of galaxies. This reconstruction is motivated by the $S_8$ tension between early-universe CMB predictions and late-time observables. We use CMB lensing data from the Atacama Cosmology Telescope DR6 and cosmic shear data from the Dark Energy Survey (DES) Y3 release to perform a gravity-only (i.e. no baryonic feedback) fit to $P(k)$ in bins of wave-number, within $\rm{\Lambda CDM}$.  We find that with DES cosmic shear data alone, $P(k)$ departs from the early-universe CMB prediction on all scales. The  joint fit with CMB lensing is consistent on large scales $k<0.2 \;{\rm Mpc}^{-1}$ but shows a $\sim 2 \sigma$ deviation from scale-independence when extending to $k = 10 \;h/\mathrm{Mpc}$. We compare our agnostic $P(k)$ reconstruction to baryonic feedback models and non-standard dark matter models: reasonable variations of both scenarios can  recover the shape and amplitude of the suppression. We discuss the advances needed to disentangle these physical effects with a full mapping of $P(k,z)$.

\end{abstract}
\maketitle

\section{Introduction}

Measurements of anisotropies in the cosmic microwave background (CMB) are remarkably well fit by the standard model of cosmology with a cosmological constant $\Lambda$ and cold dark matter (CDM) component.  While these measurements probe the recombination epoch ($t\sim 380,000$ yr), the fit to this model can be extrapolated to late times to provide predictions for the growth of structure, and for the amplitude of the matter power spectrum at the present day. In the last decade, comparisons of the predictions based on the CMB as seen by the \textit{Planck} satellite \cite{Planck_2018} have given mixed results when compared to a variety of late-time probes of the growth of structure. 

The formation of inhomogeneities in the distribution of matter can be quantified by the parameter $\sigma_8$, which is the root-mean-square variation in the matter field on scales of $8\;h/$Mpc extrapolated to the present day. A related parameter is the combination $S_8 \equiv (\Omega_m/0.3)^{0.5} \sigma_8$, chosen due to the degeneracy between the matter density $\Omega_m$ and $\sigma_8$ in  observations of gravitational weak lensing. Anisotropies in the primary CMB measured by the \textit{Planck} satellite provide a precise prediction through the \LCDM~model of $S_{8} = 0.832 \pm 0.013$ \cite{Planck_2018}.  Meanwhile, measurements of distortions in galaxy shapes (i.e. galaxy weak lensing or cosmic shear) prefer $S_8 = 0.766 \pm 0.020$, 
$S_{8}= 0.776 \pm 0.017$, and $S_8 = 0.769 \pm 0.034$ for the Kilo-Degree Survey (KiDS-1000)~\cite{Heymans2021KiDS}, Dark Energy Survey~\cite{Abbott2022DarkEnergy}, and the Hyper Suprime-Cam~\cite{2023PhRvD.108l3518L}, respectively. This $\sim 2-3\sigma$ discrepancy between measurements is known as the $S_8$ tension. A similar tension has been observed in several other astrophysical observables such as peculiar velocity measurements and the Lyman-$\alpha$ forest~\cite{Nguyen2023EvidenceFor,2025PhRvR...7a2018R,Fernandez2024CosmologicalConstraints}. Intriguingly, measurements of the lensing of the CMB are in strong agreement with the CMB prediction (e.g. \cite{Planck2018Lensing,Madhavacheril_2024,Qu_2024,Bianchini2020SPTLensing}). 

While many of the probes above disagree on their inference of $S_8$, it should be noted that they do so while probing somewhat different length scales and redshifts. The observed data is fit to a model, usually a close variant of $\Lambda$CDM, to then extrapolate the amplitude of the matter power spectrum on scales of $k\sim 1/(8\;\mathrm{Mpc}/h)$ and at the present day. For this reason, the $S_8$ tension could point to erroneous assumptions in our modeling choices. For example, the $S_8$ tension can be reduced when introducing a scale-dependent suppression in the matter power spectrum. Such solutions are numerous, and deviations from $\Lambda$CDM can arise in various ways. In the non-relativistic regime, the growth of CDM perturbations in linear theory is well described by a damped harmonic oscillator such that
\begin{align}
    \ddot{\delta} + \underbrace{2H\dot \delta}_{\rm Damping\; term} - \underbrace{4\pi G \bar{\rho} \delta}_{\rm Mass\; term} = \underbrace{0}_{\rm Source\; term},\label{eq:delta_evol}
\end{align}
where $H$ is the Hubble expansion rate, $\bar{\rho}$ is the mean matter density, and $\delta \equiv \rho/\bar{\rho}-1$ is the density contrast. Scale-dependent growth in the linear regime can arise from the modification of any of the underlined terms.

\textit{Damping term}: The rate of expansion of the Universe dampens the growth of perturbations. Late-time modifications in this rate can lead to scale-dependence in the evolution of $\delta$. For instance, kinematic couplings of dark matter to dark energy introduce such a correction to the dampening term and have been suggested as a potential solution to the $S_8$ tension~\cite{Simpson2010ScatteringOf,Poulin2023Sigma8Is,Lague2024ConstraintsOn}. A similar phenomenon occurs when dark matter couples to dark radiation~\cite{Mazoun2024ProbingInteracting}.

\textit{Mass term}: Cosmological components with a repulsive self-interaction or a non-zero pressure have a sound speed (or in some cases an effective sound speed if the pressure is due to high particle velocity or wave effects) that leads to a Jeans instability. Above a certain distance, known as the Jeans length, the repulsive forces that overcome gravitational clustering are suppressed. This manifests as a scale-dependent growth of perturbations. Many solutions to the $S_8$ tension exhibiting this behavior are particularly well-motivated by particle physics and include ultralight axions, warm dark matter, self-interacting dark matter, and atomic dark matter~\cite{Cyr_Racine_2016, Markovic_2011,Tulin_2018,2025PhRvR...7a2018R}. Since the mass term above is obtained by using Poisson's equation, modifications to the force of gravity can also introduce scale-dependence. In the case of modified gravity, the mass term can often be modeled with an effective gravitational constant that can be a function of scale~\cite{Tsujikawa2007MatterDensity}.

\textit{Source term}: The source term can be modified if there is a mechanism for dark matter to exchange energy with another component. An example of such models with regards to the $S_8$ tension is interacting dark energy~\cite{Shah2024ReconcilingS8}.

There is also the possibility of having all terms in Eq.~(\ref{eq:delta_evol}) by modified. This can happen if the overall growth of the perturbations (the time component of $\delta$) is assumed to differ from the General Relativity (GR) prediction. A useful approximation to search for time-dependent rather than scale-dependent deviations in GR is to treat the differential growth rate as a power law based on the matter density. This model is parameterized as $\frac{d\ln \delta}{d\ln a} = \Omega_m(a)^\gamma$, where the GR prediction lies at $\gamma\approx 0.55$~\cite{Wang1998ClusterAbundance,Linder2005CosmicGrowth}. Redshift-dependent deviations in the growth rate of matter perturbations have been proposed to reconcile high redshift and low redshift probes~\cite{Nguyen2023EvidenceFor}. Aside from GR extensions, these include models of dynamical dark energy and some interacting dark sectors.

Baryonic feedback processes, which are still not very well understood, have a confounding effect with observational evidence for new non-standard forms of dark matter. These affect the late-time clustering of matter on small scales; enhanced feedback (beyond what is usually considered in cosmological simulations) can alleviate the $S_8$ tension between galaxy weak lensing and the CMB~\cite{amon_2022}. Studies of the gas profiles of galaxy groups with the kinematic Sunyaev-Zel'dovich effect also favor higher levels of baryonic feedback \cite{Schaan_2021,Amodeo_2021,boryana_ksz}. A combination of X-ray, weak lensing, and kinematic Sunyaev-Zel'dovich data suggests baryonic feedback may be higher than found in simulations but not sufficiently to resolve the tension on its own~\cite{Schneider2022ConstrainingBaryons}. 

There have been several past efforts to reconstruct the matter power spectrum as a function of redshift and scale, given its usefulness in diagnosing discrepancies between data-sets beyond just the $S_8$ parameter. Among them, the matrix-based approach of Ref. \cite{Tegmark2002SeparatingThe} has been adapted to the Lyman-$\alpha$ forest~\cite{Chabanier2019MatterPower} and weak lensing~\cite{Doux2022DES}. A double-power law reconstruction of $P(k,z)$ was used in Ref. \cite{2409.13404}, which compresses all the $k$-dependence into an amplitude and tilt (but also assumes a power-law in the redshift-dependent scale factor). Ref.~\cite{amon_2022} introduced a phenomenological model that parameterizes the suppression of the matter power spectrum with a single variable $A_{\rm mod}$, and uses it to fit KiDS-1000 weak lensing data. In that model, the $A_{\rm mod}$ parameter can be physically understood as a suppression on small scales due to baryonic feedback or due to non-standard dark matter. The same model is also used to fit the Dark Energy Survey Year 3 cosmic shear data (DES-Y3 henceforth) in \cite{preston_2023}, finding a similar value for $A_{\rm mod}$ to that found for KiDS-1000. Refs. \cite{preston_2023} and \cite{preston_2024} also provide extensions of the original model to understand the redshift and scale dependence of the matter power spectrum suppression. These papers conclude that cosmic shear data from current state-of-the-art surveys (and even Stage-4 surveys) on its own does not have sufficient statistical power to help with the redshift reconstruction of this suppression. The scale dependence is found to require suppression of the matter power spectrum on scales $k< 0.3\;h$/Mpc to reconcile DES-Y3 and \textit{Planck} $\Lambda$CDM. 

In this paper, we develop a flexible phenomenological model of the matter power spectrum which allows us to test the viability of all three classes of solutions to resolve the $S_8$ tension. Moreover, we fit this model to galaxy weak lensing and CMB lensing data jointly with the aim of disentangling baryonic feedback and new physics affecting the history of structure formation. 

\section{Model}
The phenomenological model we propose here is agnostic to specific baryonic feedback or non-standard dark matter models. Instead, we aim to reconstruct the matter power spectrum as a function of scale using the parameterization

\begin{equation}\label{eq.Pk}
    P(k,z) = \sum_{i}^{N_{\alpha}} \alpha_i \mathbb{B}(k,k^l_i,k^h_i) P^{\rm fid}(k, z)
\end{equation}

where $P^{\rm fid}(k,z)$ is a fiducial total matter power spectrum (including both linear and non-linear contributions), the $\alpha_i$ parameters alter the amplitude of power over the $i$-th scale bin $[k_{i}^l,k_{i}^h]$, and the $\mathbb{B}(k)$ functions describe a top-hat function such that:

\begin{equation}
    \mathbb{B}(k,k^l_i,k^h_i) = 
    \begin{cases}
    1, & \text{if}\ k_{i}^l < k \leq k_{i}^h \\
    0, & \text{otherwise}\ 
    \end{cases}
\end{equation}

The weak lensing angular cross-power spectrum (as a function of multipole $L$) between two distributions of sources is then given by a Limber integral of the form: 
\begin{equation}\label{eq:kappaeq}
    C_{ab}^{ \kappa \kappa}(L) = \int_{0}^{z_{*}} \frac{dz}{c}\frac{H(z)}{\chi^2(z)} W_a(z) W_{b}(z) P(k=L/\chi,z)
\end{equation}
where the lensing kernel $W_{a}(z)$ or $W_{b}(z)$ are given by
\begin{equation}\label{eq:lensingkernel}
W_{a}(z) = \frac{3 \Omega_m}{2c} \frac{H_{0}^2}{H(z)}(1+z)\chi(z) \int_{z}^{\infty} dz' n_{a}(z) \frac{\chi(z')-\chi(z)}{\chi(z')}
\end{equation}

In the case of cosmic shear, $n_a(z)$ is the normalized redshift distribution of the source galaxies in a particular redshift bin. In the case of CMB lensing, this distribution is instead a $\delta$-function at the redshift of the last scattering surface, $z_{*}$:

\begin{equation}\label{cmbkernel}
W^{\mathrm{CMB}}(z) = \frac{3 \Omega_m}{2c} \frac{H_{0}^2}{H(z)}(1+z)\chi(z)\frac{\chi(z_{*})-\chi(z)}{\chi(z_{*})}
\end{equation}

The lensing convergence power spectrum for different redshift bins, $C^{\kappa \kappa}_{ab}(L)$, is in turn related to the cosmic shear correlation functions, $\xi_{\pm}$, as follows:

\begin{equation*}
 \xi_{\pm}(\theta) = \sum_{L} \frac{2L +1}{2 \pi L^2 (L+1)^2} \left[G_{L,2}^{+}(\mathrm{cos}\theta) \pm G_{L,2}^{-}(\mathrm{cos}\theta) \right]  
 \end{equation*}
 \begin{equation}\label{eq:shear}
 \times \left[ C^{\kappa \kappa ,EE}_{ab}(L) + C^{\kappa \kappa ,BB}_{ab}(L) \right] 
\end{equation}
where $\theta$ is angle on the sky and the functions $G_{L,2}^{\pm}(x)$ are functions of Legendre polynomials \cite{Secco2022CosmicShear, krause2021darkenergysurveyyear}. In this work, we reconstruct the shape of the matter power spectrum by inferring the parameters $\alpha_{i}$ in Eq.~(\ref{eq.Pk}) from observations of angular power spectra of CMB lensing and/or galaxy weak lensing shear correlation functions.

\subsection{Comparison to $A_{\rm mod}$ model and other works.} \label{subsection:Amod}

The parameterization introduced here should be contrasted with the $A_{\rm mod}$ approach from \cite{amon_2022} and \cite{preston_2023}. Those works parameterize the difference between the non-linear and the linear matter power spectrum and quantifies departures from this as a suppression as shown in Eq.~(\ref{eq:amod_model}),

\begin{equation}
    P_{m}(k,z) = P_{m}^{\rm L}(k,z) + A_{\rm mod}[P_m^{\rm {NL,NF}}(k,z) - P_{m}^{\rm L}(k,z)],
\label{eq:amod_model}
\end{equation}
where $P_m^{\rm{NL,NF}}$ refers to the non-linear matter spectrum due to cold-dark matter contributions only (ignoring baryonic feedback), $P_m^{\rm L}$ is the linear matter power spectrum, and the parameter $A_{\rm mod}$ can be physically interpreted as a suppression due to baryonic feedback or non-standard forms of dark matter. Ref.~\cite{preston_2023} proposed a similar model to Ref.~\cite{amon_2022} but accounting for scale-dependent suppression by replacing $A_{\rm mod}$ with five ${A_{i}}$ parameters that act on $P_m(k,z)$ in different $k$-bins.

If we restrict $P(k_i)$ in our model in Eq.~(\ref{eq.Pk}) to only account for cold-dark matter contributions to the matter power spectrum and ignoring baryonic feedback, then our model and the model in Ref.~\cite{preston_2023} are equivalent as follows:
\begin{equation}\label{eq:amod}
    \sum_{i}^{N_{\alpha}} \alpha_i P_{m}^{\rm{NL,NF}}(k_i) = P_m^{\rm{L}}(k) + \sum_{i}^{N_{A}} A_{i}[P_m^{\rm{NL,NF}}(k_{i}) - P_{m}^{\rm{L}}(k_{i})].
\end{equation}
Furthermore, if both models have the same $k$-binning, then
\begin{equation}
    \alpha_{i} = \frac{P_m^{\rm{L}}(k_i)}{P_{m}^{\rm{NL,NF}}(k_i)} + A_{i} \left( 1 - \frac{P_m^{\rm{L}}(k_i)}{P_{m}^{\rm{NL,NF}}(k_i)} \right).
\end{equation}

Although here we made a direct comparison between our model and that of ~\cite{amon_2022,preston_2023}, we point out that different analysis choices were made in the present work than in those works, as we detail in sections \ref{section:data} and \ref{section:methods}. Importantly, our parameterization is better suited for assessing whether there is a {\it scale-dependent} departure from \LCDM. For example, if observations were uniformly discrepant from $\Lambda$CDM by an amplitude on all scales (say, due to a multiplicative bias in galaxy weak lensing), the $\Amod$ parameterization in Eq.~(\ref{eq:amod}) would indicate an incorrectly large deviation on small-scales, since it does not allow the large scales to disagree with \Planck. The binned $\Amod$ parameterization with multiple $A_i$ parameters in  \cite{preston_2023} does not fully address this since the prior ranges used there are $U[0.5,1]$; i.e., the inferred parameters will be biased low since only suppression is allowed, and not enhancement.

More recently, \cite{ye2024} also introduced a model-independent reconstruction of the matter power spectrum. This model performs a perturbative expansion of the linear growth factor, and fits cosmological data at various redshifts (galaxy weak lensing, CMB lensing, BAO and supernovae) to obtain coefficients that scale the linear power spectrum as a function of scale. Their model, which attempts to reconstruct $P(k)$ to very high $k$ with only a first-order expansion, recovers the linear power spectrum at large scales, while at small scales, it predicts some non-linear contributions to the matter power spectrum without assuming any specific non-linear model. In particular, at small scales ($ k > 1 \;\mathrm{Mpc}^{-1}$), the model seems to predict an enhancement of the matter power spectrum. Our approach is significantly different since it performs a direct reconstruction of the full non-linear matter power.

\section{Data} \label{section:data}

Unlike previous work~\cite{Tegmark2002SeparatingThe,Chabanier2019MatterPower,Doux2022DES,ye2024},  we perform a joint fit of  both CMB lensing and galaxy weak lensing data to directly reconstruct the non-linear matter power spectrum. To this end, we applied the model described above to lensing data from the Atacama Cosmology Telescope (ACT) Data Release 6 (ACT-DR6 henceforth) and to DES-Y3. The ACT-DR6 lensing power spectrum is obtained from night-time CMB data acquired by ACT between 2017 and 2022, and covers roughly 9400 deg${}^2$. Details on the processing of the ACT DR6 data can be found in \cite{Qu_2024}. In this paper in particular, we use the ACT+\textit{Planck} lensing power spectrum with the extended multipole range ($L<1250$). We include this data-set by using the publicly available likelihood used to constrain cosmology from ACT DR6 lensing data, but ran this in the ``Limber integral mode''. The Limber approximation is precise for the scales we use ($L>40$).  Further details on this likelihood can be found in \cite{Madhavacheril_2024}.
For our galaxy weak lensing dataset, we use the DES-Y3 cosmic shear likelihood as implemented in \texttt{CocoA} (as introduced in \cite{Zhong2023GrowthAnd}), which uses the CosmoLike framework \cite{Krause2017cosmolike}. The DES-Y3 data was collected over the first three years of the survey and covers about  4143 deg${}^2$. Details about the data and the analysis can be found in ~\cite{DESY3_cosmicshear_amon}.

\section{Methodology}\label{section:methods}

\begin{figure*}
    \includegraphics[width=1.0 \textwidth]{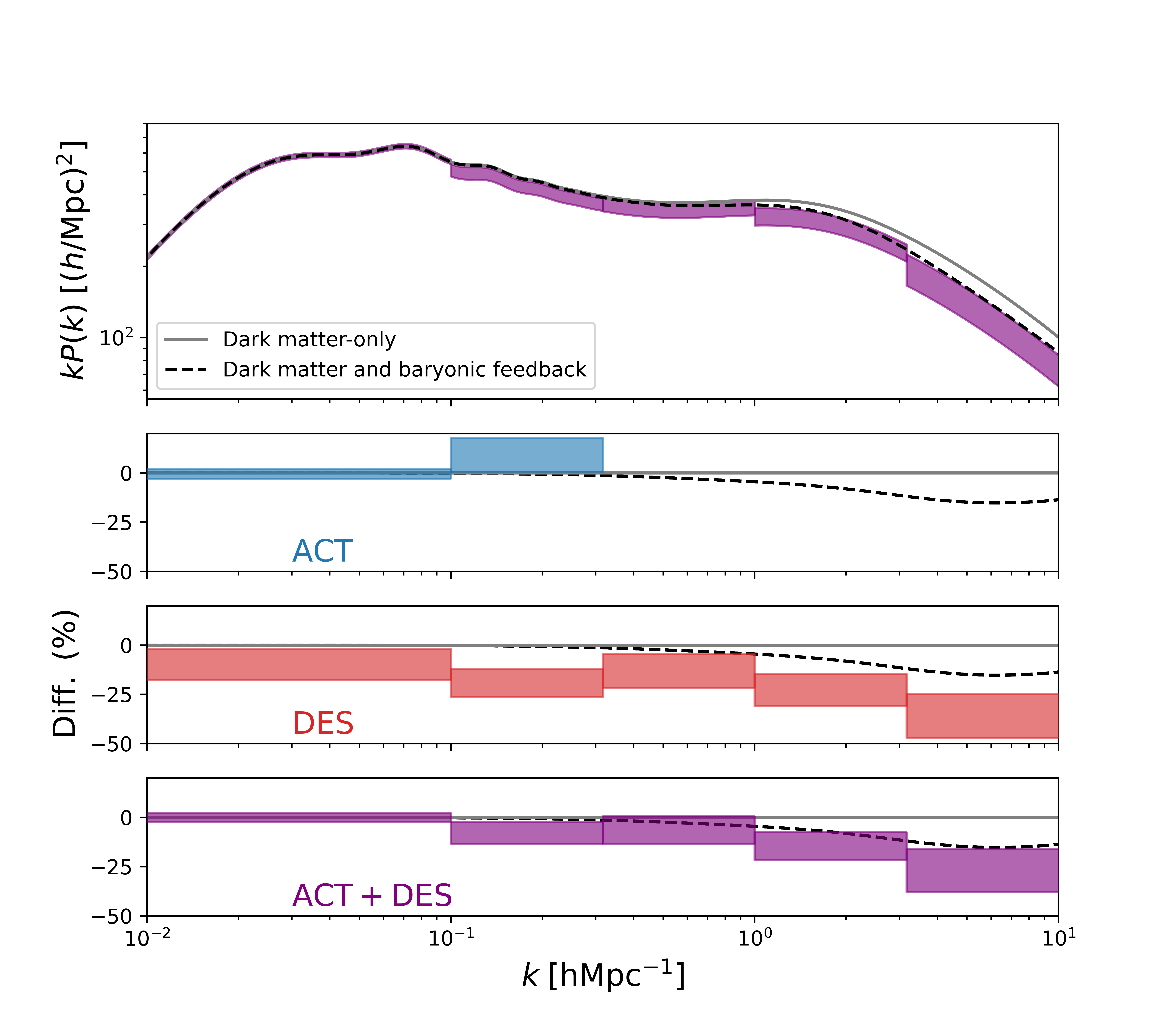}
    \caption{The non-linear matter power spectrum (at redshift $z=0$) as a function of wavenumber $k$, measured through fitting deviations from a gravity-only extrapolation of \Planck~CMB fits with the \texttt{Mead2020} non-linear model. In the top panel, we show a fit (purple) that uses both ACT CMB lensing and DES galaxy weak lensing jointly. Differences with respect to the baseline model are also shown for ACT-alone (blue) and DES-alone (red).  We also show curves corresponding to the \texttt{Mead2020-feedback} non-linear power spectrum prediction that includes baryonic feedback, with $\log{T_{\rm AGN}} = 7.8$.}
    \label{fig:Pk_nonlin}
\end{figure*}

Since our approach gives considerable freedom to the shape of the matter power spectrum at $z=0$, we may proceed by fixing the cosmological parameters; we choose as our fiducial parameters determined from TT,TE,EE+lowE in the \textit{Planck} 2018 CMB  analysis \cite{Planck_2018} (Table \ref{table:Planck_cosmo}). It should be noted that this does fix the redshift evolution of background and perturbation quantities to be that determined by this cosmology. We modified the Boltzmann code CAMB \cite{CAMB} to introduce the $\alpha_i$ parameters that reshape the non-linear matter power spectrum as a function of scale, with a binning set by

\begin{equation} \label{eq:Pkbins}
    \begin{cases}
    \alpha_{1}, & \text{if}\ k ~\left[h/\mathrm{Mpc}\right] \leq 0.10 \\
    \alpha_{2}, & \text{if}\ 0.10  < k ~\left[h/\mathrm{Mpc}\right] \leq 0.32 \\
    \alpha_{3}, & \text{if}\ 0.32  < k  ~\left[h/\mathrm{Mpc}\right] \leq 1.00  \\
    \alpha_{4}, & \text{if}\ 1.00  < k ~\left[h/\mathrm{Mpc}\right] \leq 3.16  \\
    \alpha_{5}, & \text{if}\ k ~\left[h/\mathrm{Mpc}\right] > 3.16 
    \end{cases}
\end{equation}

\begin{center}
\begin{table}
\begin{tabular}{ c c  } 
 Parameter & TT,TE,EE+lowE value \\ 
 \hline
$\Omega_b h^2$ & 0.02236 \\  
$\Omega_c h^2$ & 0.1202 \\ 
$\tau$ & 0.0544 \\ 
$\ln(10^{10}A_s)$ & 3.045  \\
$n_s$ & 0.9649 \\
$H_0$ & 67.27  \\
\end{tabular}
\caption{TT,TE,EE+lowE \textit{Planck} 2018 parameters used as the fiducial cosmology in this paper.}
\label{table:Planck_cosmo}
\end{table}
\end{center}

We performed five different fits, which we summarize in Table \ref{table:variants}. We run three of the fits with the nonlinear model \texttt{Mead2020} in \texttt{CAMB} \cite{Mead_2020}, which only accounts for dark matter contributions when calculating the non-linear matter power spectrum, and the two others with \texttt{Mead2020-feedback}, which also includes baryonic feedback contributions. We set the feedback parameter $\log{T_{\rm AGN}} = 7.8$  for the runs with the \texttt{Mead2020-feedback} model. 

For the cases without baryonic feedback, we either fit ACT-DR6 CMB lensing and DES-Y3 cosmic shear data  likelihoods jointly, or fit each data set separately. For the ACT-DR6-only run, we only fit for $[\alpha_1,\alpha_2]$ since we expect CMB lensing data to only accurately constrain the largest scales. The DES-Y3 cosmic shear-only run and the joint run do fit for the five $\alpha_i$ parameters.
For the cases with baryonic feedback, we also do a joint ACT-DR6 lensing and DES-Y3 cosmic shear data fit and a DES-only fit.
We note that we have used the entire cosmic shear correlation data without applying the scale cuts imposed in the original DES-Y3 cosmic shear analysis \cite{DESY3_cosmicshear_amon}. This is because applying the conservative scale cuts from the DES-Y3 analysis would prevent us from reconstructing the matter power spectrum on small scales.

For any of the fits described above that make use of DES-Y3 cosmic shear, we also sampled over the DES-Y3 nuisance parameters described in Table \ref{table:des_nuisance}, where $N(\mu, \sigma)$ represents a Gaussian prior with mean $\mu$ and variance $\sigma^2$, and where $U(a, b)$ represents a uniform distribution over the interval $(a, b)$. These priors were taken from the official DES-Y3 cosmic shear analysis \cite{DESY3_cosmicshear_amon}. As in the fiducial DES-Y3 cosmic shear analysis, we employ the Tidal Alignment and Tidal Torquing model (TATT), which consists of 5 parameters, $a_1,a_2,\eta_i,\eta_2$ \cite{Blazek_2019}. The priors for the $\alpha_{i}$ parameters were set to a uniform distribution $U(0.3,1.7)$.

\begin{center}
\begin{table*}
\begin{tabular}{ c c c c c} 
 Label & Data and likelihood & Halofit model & Sampled parameters \\ 
 \hline
 ACT & ACT-DR6 & \texttt{Mead2020} & $\alpha_1$, $\alpha_2$\\
 DES & DES-Y3-CS & \texttt{Mead2020} & $\alpha_1$,..,$\alpha_5$, nuisance \\ 
 ACT+DES & ACT-DR6 + DES-Y3-CS & \texttt{Mead2020} & $\alpha_1$,..,$\alpha_5$, nuisance \\ 
 DES-BF & DES-Y3-CS & \texttt{Mead2020-feedback} & $\alpha_1$,..,$\alpha_5$, nuisance \\ 
 ACT+DES-BF & ACT-DR6 + DES-Y3-CS & \texttt{Mead2020-feedback} & $\alpha_1$,..,$\alpha_5$, nuisance \\ 
\end{tabular}
\caption{Description of the different fits, the parameters sampled and the data involved in the fit. ACT-DR6 refers to ACT-DR6 lensing, while DES-Y3-CS to DES-Y3 cosmic shear. Nuisance refers to the DES nuisance parameters described in Table \ref{table:des_nuisance}. The -BF fits are with respect to a model with moderate baryonic feedback (\texttt{Mead2020-feedback}) while all others are with respect to the gravity-only non-linear model (\texttt{Mead2020}).}
\label{table:variants}
\end{table*}
\end{center}

\begin{center}
\begin{table} 
\begin{tabular}{ c c}
Parameter & Prior  \\
\hline
\textbf{Source photo-z} &  \\
$\Delta z_{s}^{1}$ & $N(0.0,0.018)$ \\
$\Delta z_{s}^{2}$ & $N(0.0,0.015)$ \\
$\Delta z_{s}^{3}$ & $N(0.0,0.011)$ \\
$\Delta z_{s}^{4}$ & $N(0.0,0.017)$ \\
\textbf{Shear calibration} &  \\
$m^{1}$ & $N(-0.006,0.009)$ \\
$m^{2}$ & $N(-0.02,0.008)$ \\
$m^{3}$ & $N(-0.024,0.008)$ \\
$m^{4}$ & $N(-0.037,0.008)$ \\
\textbf{Intrinsic alignment} & \\
$a_{i}, i \in \{1,2\}$ & $U(-5.0, 5.0)$ \\
$\eta_{i}, i \in \{1,2\}$ & $U(-5.0, 5.0)$ \\
$b_{TA}$ & $U(0,2)$ \\
\end{tabular}
\caption{DES-Y3 nuisance parameters that are marginalized over when using DES-Y3 cosmic shear data.}
\label{table:des_nuisance}
\end{table}
\end{center}

\section{Results}
Fig.~\ref{fig:Pk_nonlin} shows the scale-dependent reconstruction of the matter power spectrum obtained from the runs with \texttt{Mead2020}. The corresponding triangle plot and correlation matrices of the parameters can be found in Appendix \ref{section:app}.

\begin{figure*}
    \includegraphics[width=1.0 \textwidth]{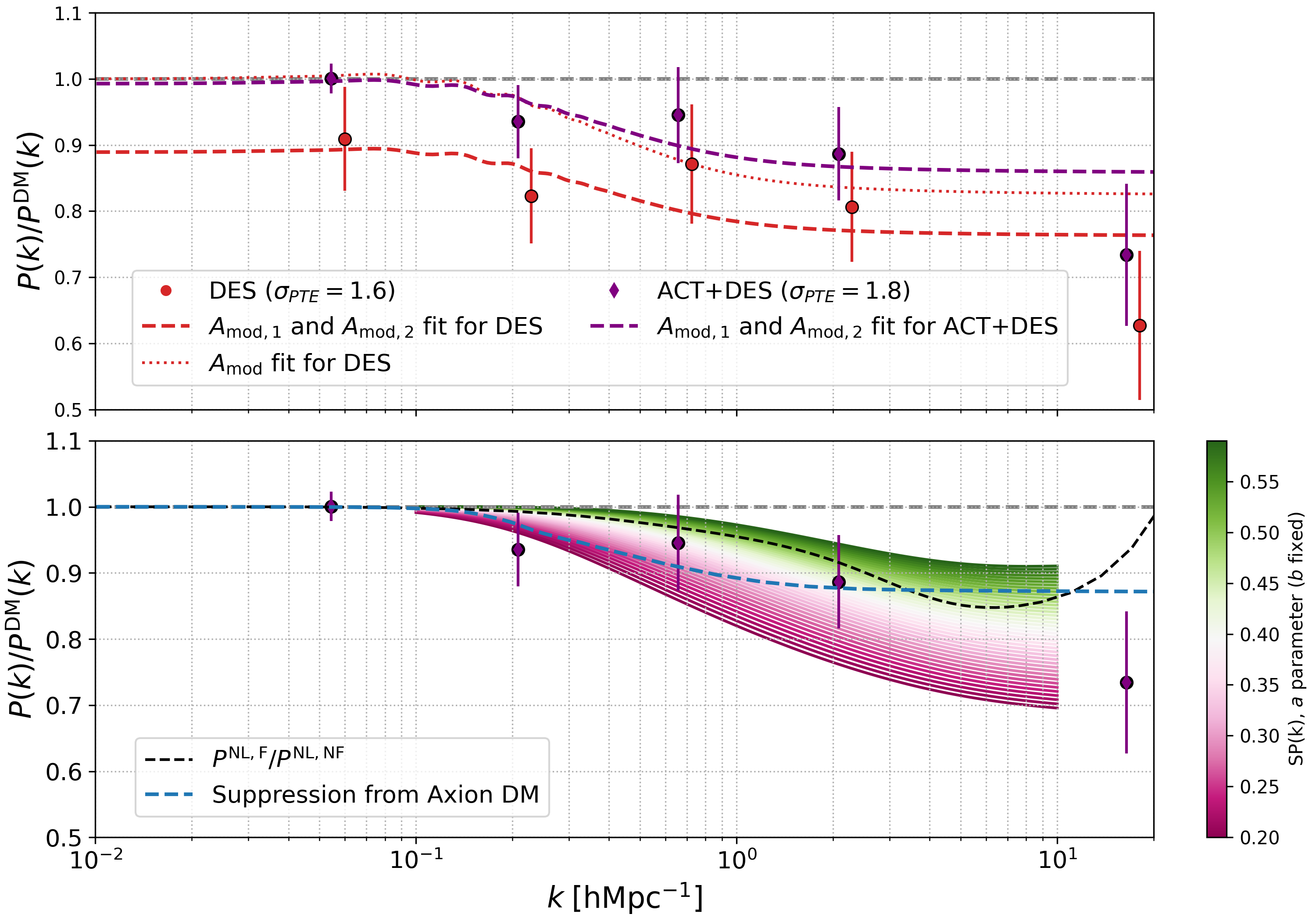}
    \caption{Inferred deviation of the non-linear matter power spectrum from our gravity-only baseline shown against various models. \textit{Top}: We show the $A_{\rm{mod}}$ model (Eq. \ref{eq:amod_model}) best-fit of the DES-alone $\alpha_{i}$ parameters in a red-dotted line, along with a variation with two $A_{\mathrm{mod},i}$ parameters (Eq. \ref{eq:amod_model_variation}), which includes an extra parameter capturing an overall amplitude deviation on all scales. We show the best-fit of this model to DES-alone (red dashed line) and ACT+DES (purple dashed line). \textit{Bottom}: We show the ratio of the non-linear matter power {\it with} baryonic feedback contributions (\texttt{Mead2020-feedback}; $\log{T_{\rm AGN}} = 7.8$) to the non-linear matter power spectrum from gravity-only (\texttt{Mead2020}) with a dashed black line. We also qualitatively compare our power spectrum reconstruction to variations of the SP(k) model for baryonic feedback  \cite{Salcido_2023} for different values of the parameter $a$ (Eq. \ref{eq:fbMhalo}) with $b=0.3$ at $z=0.125$ (pink-to-green gradient lines). The SP(k) model was only calibrated up to $k\sim 10 \ h\mathrm{Mpc}^{-1}$, hence the abrupt cut. Finally, we include the best-fit ultralight axion dark matter model from \cite{2025PhRvR...7a2018R} for comparison in blue.}
    \label{fig:alphas_pte}
\end{figure*}

In order to quantify the amount of scale dependence that these $\alpha_i$ parameters imprint on the matter power spectrum, we treat the best-fit for the $\alpha_{i}$ parameters as data points (with covariance obtained from the MCMC samples for the parameters) and then fit these to a model consisting of just a constant offset. A good fit to this constant model would  be an indication of scale-independence, i.e. a disagreement with the \Planck~prediction would then be due to an overall amplitude shift. We then compute the $\chi^2$ and probability-to-exceed (PTE) for these fits, noting that a smaller PTE corresponds to a more significant scale-dependent departure from the \Planck~prediction. Table \ref{table:pte} and Fig.~\ref{fig:alphas_pte} summarize the PTE for the different fits. Note that we either fit for all the available $\alpha_{i}$ parameters ($\alpha_{1},...,\alpha_{5}$), or leave out the last bin ($\alpha_{1},...,\alpha_{4}$), which corresponds to the smallest scales. We notice that the case for a scale-dependent departure from $\Lambda \mathrm{CDM}$ is stronger for the five $\alpha_i$ parameters obtained from the joint ACT+DES fit with \texttt{Mead2020}. When accounting for some amount of baryonic feedback, as we do in the runs with  \texttt{Mead2020-feedback}, or alternatively, when leaving out the smallest scales from the fit, the evidence for scale-dependence is much weaker. 

\begin{center}
    \begin{table}
        \begin{tabular}{c c c c c}
        Label & Parameters fit & Amplitude & PTE \\
        \hline
        ACT+DES & $\alpha_{1},...,\alpha_{5}$ & $0.98 \pm  0.02$ &
        0.07 \\
        DES & $\alpha_{1},...,\alpha_{5}$ & $0.85 \pm 0.05$ & 0.12 \\
        ACT+DES & $\alpha_{1},...,\alpha_{4}$ & $0.98 \pm 0.02 $ & 0.40 \\ 
        DES & $\alpha_{1},...,\alpha_{4}$& $0.86 \pm 0.05$ & 0.79 \\
        ACT+DES-BF & $\alpha_{1},...,\alpha_{5}$ & $0.98 \pm 0.02$ & 0.20 \\
        DES-BF & $\alpha_{1},...,\alpha_{5}$ & $0.92 \pm 0.05$ & 0.25 \\
        \end{tabular}
        \caption{Constant amplitude fits to the $\alpha_{i}$ parameters for the various data combinations. The -BF fits are with respect to a model with moderate baryonic feedback (\texttt{Mead2020-feedback}) while all others are with respect to the gravity-only non-linear model (\texttt{Mead2020}).}
        \label{table:pte}
    \end{table}
\end{center}

We also fit a slight variation of the model introduced in \cite{amon_2022}, where we introduce an additional $A_{\rm mod,1}$ parameter that frees up the amplitude of the matter power spectrum:

\begin{equation}
  \label{eq:amod_model_variation}
    P_{m}(k,z) = A_{\rm mod,1}P_{m}^{\rm{L}}(k,z) + A_{\rm mod,2}[P_m^{\rm NL}(k,z) - P_{m}^{\rm{L}}(k,z)]
\end{equation}

\begin{center}
    \begin{table}
        \begin{tabular}{c c c c}
        Label & $A_{\rm mod,1}$ & $A_{\rm mod,2}$ & PTE\\
        \hline
        ACT+DES & $0.99\pm0.02$ & $0.86\pm0.05$ & 0.50 \\
        DES & $0.89\pm0.05$ & $0.76\pm0.07$ & 0.34 \\
        \hline
        ACT+DES & $1.0$ & $0.86\pm0.05$ & 0.64 \\     
        DES & $1.0$ & $0.82\pm0.06$ & 0.11 \\
        \end{tabular}
        \caption{Results from fitting a variation of the model in \cite{amon_2022} with two $A_{\rm mod}$ parameters to the $\alpha_{i}$ parameters, as described in Eq. \ref{eq:amod_model_variation}. 
        The third and fourth rows are the fit to the original model where, by definition, $A_{\rm{mod},1} = 1$. }
        \label{table:two_amods_table}
    \end{table}
\end{center}

For comparison with \cite{amon_2022}, we also fit the original $A_{\rm mod}$ variant, which corresponds to fixing $A_{\rm mod,1}=1$. The result of these fits is shown in Fig. \ref{fig:alphas_pte}, and summarized in Table \ref{table:two_amods_table}. In the case of the DES-only runs, we calculate the Akaike Information Criterion (AIC = $2k+\chi^2$) to compare the $A_{\rm mod}$ and $A_{\rm mod,1}+A_{\rm mod,2}$ models and find $\Delta{\rm AIC}=1.1$ suggesting that one model is not significantly preferred over another. However, freeing up the large-scale $A_{\rm mod,1}$  parameter for DES-only does give $A_{\rm mod,1}=0.89\pm0.05$, a $2.2\sigma$ deficit compared to the \Planck~ prediction.  This deficit disappears ($A_{\rm mod,1}=0.99\pm0.02$) in the joint ACT+DES run.  

Although the results obtained with the halofit model \texttt{Mead2020} in this paper are agnostic to specific baryonic feedback models or to non-standard dark matter theories, we also compare the reconstruction of the matter power spectrum in this work to interesting models. 
For example, in Fig. 2 we also we also overplot SP(k) hydrodynamical simulation-based models for the suppression of the non-linear matter power spectrum due to baryonic feedback from \cite{Salcido_2023}. 
These models employed 400 simulations of an ample range of baryonic feedback processes at a wide range of scales and redshifts, and mapped the suppression to the total baryon fraction of haloes, $f_{b}$, as  follows: 

\begin{equation} \label{eq:SPK}
\frac{P_{\rm hydro}}{P_{\rm DM}} = \lambda(k,z) - \left(\lambda(k,z) - \mu(k,z) \right) \exp{(-v(k,z)\tilde{f_{b}})},
\end{equation}
where $\tilde{f_{b}}$ is the baryon fraction at some optimal halo mass: $\tilde{f_{b}} = f_{b}(\tilde{M}_{k,200c}(k,z))/(\Omega_b/\Omega_{m})$. Details on this parameter and on $\lambda(k,z)$, $\mu(k,z)$ and $v(k,z)$, as well as their corresponding best-fit parameters can be found in ~\cite{Salcido_2023}. With these best-fit parameters, and the functional form for the $f_{b} - M_{\rm halo}$ relation defined as

\begin{equation} \label{eq:fbMhalo}
    f_{b} / (\Omega_b/\Omega_m) = a \left( \frac{M_{200c}}{10^{13.5}M_{\odot}} \right)^{b}
\end{equation}
SP(k) can return predictions for the amount of suppression to the non-linear matter power spectrum due to baryonic feedback down to redshift z=0.125 and for $b=0.3$, which we overplot in Fig.~\ref{fig:alphas_pte}. At first glance, the range of curves for this model  adequately recovers  the shape and amplitude described by the $\alpha_{i}$ parameters.

We additionally overplot in Fig.~\ref{fig:alphas_pte} a prediction for the suppression of the non-linear matter power spectrum due to ultralight axion dark matter. We use the best-fit model in ~\cite{2025PhRvR...7a2018R} that alleviates the tension arising in the slope of the linear matter power spectrum from the Lyman-$\alpha$ forest and CMB, which was found to be an axion with mass $10^{-25.6}$ eV that constitutes $1.3$\% of the dark matter (with the rest being cold dark matter). This value is well within the constraints on axions from CMB and galaxy surveys~\cite{Lague2022ConstrainingUltralight,Rogers2023UltralightAxions}. The exact shape of the suppression is obtained from \texttt{axionHMcode}~\cite{Vogt2023ImprovedMixed} which computes the non-linear corrections to the matter power spectrum calculated by the adapted Boltzmann code \texttt{axionCAMB}~\cite{Hlozek2015ASearch}. We notice that this suppression is similar to the best-fit of the two $A_{\mathrm{mod},i}$ model we fit to the $\alpha_{i}$ parameters from the joint ACT+DES run, and could adequately describe the shape and amplitude of the $\alpha_{i}$ parameters, as seen in Fig. \ref{fig:alphas_pte}.

\section{Conclusion}

The results presented here constitute the first attempt at directly reconstructing the non-linear matter power spectrum across scales with both CMB lensing and cosmic shear data. When ignoring baryonic feedback, we find a mild indication of a scale-dependent suppression, as evidenced by the joint fit with ACT DR6 lensing and DES-Y3 cosmic shear runs in Fig \ref{fig:Pk_nonlin}. However, either of these data-sets on their own do not indicate such a preference, in contrast to what was found in previous work \cite{amon_2022,preston_2023}. We also perform a reconstruction with respect to a model that includes a moderate amount of baryonic feedback (the default in \texttt{Mead2020-feedback}): here, the model provides a good fit with no suggestion of a scale-dependent departure even when jointly fit to both CMB lensing and weak lensing. Note that our analysis assumes the standard \LCDM~cosmology (except for the one dark matter alternative discussed below) with the parameters listed in Table 1 fixed to the values determined by the \textit{Planck} 2018 primary CMB analysis. 

A comparison with the model in \cite{amon_2022,preston_2023} and their best-fit parameter from cosmic shear data ($A_{\rm mod}=0.8$) shows that such a model roughly describes the shape and the scale of the suppression recovered with the {\it joint} ACT+DES fit of the $\alpha_{i}$ parameters in this paper, in particular for the two central wavenumber bins. However, this does not seem to be the case when using only DES galaxy weak lensing data.  We introduced a  variation of this model, where two $A_{\mathrm{mod},i}$ parameters are introduced, one of which ($A_{\mathrm{mod},1}$) also varies the amplitude of the linear matter power spectrum. Particularly in the case of DES data alone, we find a $2.2\sigma$ preference for $A_{\mathrm{mod},1}<1$ (although we note that the two-parameter model is not a better fit to the data than the standard one-parameter $A_{\mathrm{mod}}$), i.e. that the DES data differs from the \Planck~ prediction on linear scales, as is also evident from the top panel of Fig. \ref{fig:alphas_pte}.

It should be noted that the central value of $\Omega_m$ found by cosmic shear measurements also differs slightly from our fiducial value based on CMB measurements. Indeed, DES-Y3 cosmic shear finds $\Omega_m=0.290^{+0.039}_
{-0.063}$~\cite{Secco2022CosmicShear} which is lower than the fiducial $\Omega_m=0.315$. Given the degeneracy direction of $\sigma_8-\Omega_m$, one needs to decrease $\sigma_8$ to obtain the same $\chi^2$ with a higher matter density when fitting cosmic shear data. This may be partially responsible for the overall (scale-independent) suppression observed in the DES-only analysis.

We also compared the $\alpha_i$ parameters recovered from  ACT+DES and the DES fit to the SP(k) model for baryonic feedback suppression of the non-linear matter power spectrum \cite{Salcido_2023}. While we do not attempt to quantitatively fit this baryonic feedback model, some variations overplotted in Fig. \ref{fig:alphas_pte} do appear recover the shape and amplitude of the suppression induced by the $\alpha_{i}$ parameters. Further work can be done to fit for the $\alpha_{i}$ parameters along with SP(k) model using both cosmic shear and CMB lensing data. 

We have also compared our matter power spectrum reconstruction to the best-fit model of ultralight axion dark matter suppression of the non-linear matter power spectrum in \cite{2025PhRvR...7a2018R}. These comparisons motivate the need for more accurate models of baryonic feedback, as it is almost impossible to distinguish between its impact on the non-linear matter power spectrum and the effects of non-standard dark matter models. Our results suggest that strong constraints on baryonic feedback models -- likely from complementary measurements such as the thermal (e.g. \cite{2108.01600,2108.01601,2109.04458,2301.02186}) and kinematic Sunyaev-Zel'dovich effects (e.g. \cite{1705.05881,1903.04647,2009.05557,2101.08373,2407.07152,2412.03631}) -- are required to be able to test dark matter models with lensing data.

Another effect of the coupling of scale and redshift dependence may be at play in  cross-correlation studies between galaxy surveys and CMB lensing \cite{White_2022,sailer2024,kim_2024,Farren_2024,karim2024desi} that measure the growth of structure over time and find a somewhat lower value of $S_8$ at lower redshifts. 
Given the form of the line-of-sight projection integral (Eq.~\ref{eq:kappaeq}), a given $L$ mode receives contributions from the matter power spectrum at smaller scales from low redshifts. Therefore a redshift-dependent suppression effect could manifest as scale-dependence in our analysis. We leave the study of time-dependent deviations from $\Lambda$CDM in the matter power spectrum to future work.  With improved galaxy weak lensing (e.g. \cite{10.48550/arXiv.1503.03757,1809.01669,2405.13491} ) and CMB lensing (e.g. \cite{1610.02743,1808.07445,2024arXiv240512220M}) measurements from upcoming or planned surveys, it may in the future be possible to  separate out redshift and scale-dependent effects, allowing the mapping of $P(k,z)$ in a grid across scales and cosmic time.

\section{Acknowledgments}
We thank Gary Bernstein, Cyrille Doux, Rafael Gomes, Tanvi Karwal, Sanjit Kobla, Meng-Xiang Lin and Kunhao Zhong for helpful discussions. We thank Neelima Sehgal and Ed Wollack for helpful suggestions for this manuscript. We are especially grateful to Vivian Miranda for help with the \texttt{Cocoa} likelihood framework.  KPS and MM acknowledge support from NSF grants AST-2307727 and  AST-2153201. AL acknowledges support from NASA grant 21-ATP21-0145.  BJ is partially supported by the US Department of Energy grant DE-SC0007901. We also acknowledge use of the \texttt{matplotlib}~\citep{Hunter:2007} package for producing plots in this paper, use of the Boltzmann code \texttt{CAMB}~\citep{CAMB} for calculating theory spectra, and use of the \texttt{GetDist} \citep{1910.13970} and \texttt{Cobaya} \citep{2005.05290} software for likelihood analysis and sampling.
\bibliographystyle{apsrev.bst}
\bibliography{biblio}

\begin{thebibliography}{71}
\expandafter\ifx\csname natexlab\endcsname\relax\def\natexlab#1{#1}\fi
\expandafter\ifx\csname bibnamefont\endcsname\relax
  \def\bibnamefont#1{#1}\fi
\expandafter\ifx\csname bibfnamefont\endcsname\relax
  \def\bibfnamefont#1{#1}\fi
\expandafter\ifx\csname citenamefont\endcsname\relax
  \def\citenamefont#1{#1}\fi
\expandafter\ifx\csname url\endcsname\relax
  \def\url#1{\texttt{#1}}\fi
\expandafter\ifx\csname urlprefix\endcsname\relax\def\urlprefix{URL }\fi
\providecommand{\bibinfo}[2]{#2}
\providecommand{\eprint}[2][]{\url{#2}}

\bibitem[{\citenamefont{{Planck Collaboration} et~al.}(2020{\natexlab{a}})\citenamefont{{Planck Collaboration}, {Aghanim}, {Akrami}, {Ashdown}, {Aumont}, {Baccigalupi}, {Ballardini}, {Banday}, {Barreiro}, {Bartolo} et~al.}}]{Planck_2018}
\bibinfo{author}{\bibnamefont{{Planck Collaboration}}}, \bibinfo{author}{\bibfnamefont{N.}~\bibnamefont{{Aghanim}}}, \bibinfo{author}{\bibfnamefont{Y.}~\bibnamefont{{Akrami}}}, \bibinfo{author}{\bibfnamefont{M.}~\bibnamefont{{Ashdown}}}, \bibinfo{author}{\bibfnamefont{J.}~\bibnamefont{{Aumont}}}, \bibinfo{author}{\bibfnamefont{C.}~\bibnamefont{{Baccigalupi}}}, \bibinfo{author}{\bibfnamefont{M.}~\bibnamefont{{Ballardini}}}, \bibinfo{author}{\bibfnamefont{A.~J.} \bibnamefont{{Banday}}}, \bibinfo{author}{\bibfnamefont{R.~B.} \bibnamefont{{Barreiro}}}, \bibinfo{author}{\bibfnamefont{N.}~\bibnamefont{{Bartolo}}}, \bibnamefont{et~al.}, \bibinfo{journal}{\aap} \textbf{\bibinfo{volume}{641}}, \bibinfo{eid}{A6} (\bibinfo{year}{2020}{\natexlab{a}}), \eprint{1807.06209}.

\bibitem[{\citenamefont{{Heymans} et~al.}(2021)\citenamefont{{Heymans}, {Tr{\"o}ster}, {Asgari}, {Blake}, {Hildebrandt}, {Joachimi}, {Kuijken}, {Lin}, {S{\'a}nchez}, {van den Busch} et~al.}}]{Heymans2021KiDS}
\bibinfo{author}{\bibfnamefont{C.}~\bibnamefont{{Heymans}}}, \bibinfo{author}{\bibfnamefont{T.}~\bibnamefont{{Tr{\"o}ster}}}, \bibinfo{author}{\bibfnamefont{M.}~\bibnamefont{{Asgari}}}, \bibinfo{author}{\bibfnamefont{C.}~\bibnamefont{{Blake}}}, \bibinfo{author}{\bibfnamefont{H.}~\bibnamefont{{Hildebrandt}}}, \bibinfo{author}{\bibfnamefont{B.}~\bibnamefont{{Joachimi}}}, \bibinfo{author}{\bibfnamefont{K.}~\bibnamefont{{Kuijken}}}, \bibinfo{author}{\bibfnamefont{C.-A.} \bibnamefont{{Lin}}}, \bibinfo{author}{\bibfnamefont{A.~G.} \bibnamefont{{S{\'a}nchez}}}, \bibinfo{author}{\bibfnamefont{J.~L.} \bibnamefont{{van den Busch}}}, \bibnamefont{et~al.}, \bibinfo{journal}{\aap} \textbf{\bibinfo{volume}{646}}, \bibinfo{eid}{A140} (\bibinfo{year}{2021}), \eprint{2007.15632}.

\bibitem[{\citenamefont{{Abbott} et~al.}(2022)\citenamefont{{Abbott}, {Aguena}, {Alarcon}, {Allam}, {Alves}, {Amon}, {Andrade-Oliveira}, {Annis}, {Avila}, {Bacon} et~al.}}]{Abbott2022DarkEnergy}
\bibinfo{author}{\bibfnamefont{T.~M.~C.} \bibnamefont{{Abbott}}}, \bibinfo{author}{\bibfnamefont{M.}~\bibnamefont{{Aguena}}}, \bibinfo{author}{\bibfnamefont{A.}~\bibnamefont{{Alarcon}}}, \bibinfo{author}{\bibfnamefont{S.}~\bibnamefont{{Allam}}}, \bibinfo{author}{\bibfnamefont{O.}~\bibnamefont{{Alves}}}, \bibinfo{author}{\bibfnamefont{A.}~\bibnamefont{{Amon}}}, \bibinfo{author}{\bibfnamefont{F.}~\bibnamefont{{Andrade-Oliveira}}}, \bibinfo{author}{\bibfnamefont{J.}~\bibnamefont{{Annis}}}, \bibinfo{author}{\bibfnamefont{S.}~\bibnamefont{{Avila}}}, \bibinfo{author}{\bibfnamefont{D.}~\bibnamefont{{Bacon}}}, \bibnamefont{et~al.}, \bibinfo{journal}{\prd} \textbf{\bibinfo{volume}{105}}, \bibinfo{eid}{023520} (\bibinfo{year}{2022}), \eprint{2105.13549}.

\bibitem[{\citenamefont{{Li} et~al.}(2023)\citenamefont{{Li}, {Zhang}, {Sugiyama}, {Dalal}, {Terasawa}, {Rau}, {Mandelbaum}, {Takada}, {More}, {Strauss} et~al.}}]{2023PhRvD.108l3518L}
\bibinfo{author}{\bibfnamefont{X.}~\bibnamefont{{Li}}}, \bibinfo{author}{\bibfnamefont{T.}~\bibnamefont{{Zhang}}}, \bibinfo{author}{\bibfnamefont{S.}~\bibnamefont{{Sugiyama}}}, \bibinfo{author}{\bibfnamefont{R.}~\bibnamefont{{Dalal}}}, \bibinfo{author}{\bibfnamefont{R.}~\bibnamefont{{Terasawa}}}, \bibinfo{author}{\bibfnamefont{M.~M.} \bibnamefont{{Rau}}}, \bibinfo{author}{\bibfnamefont{R.}~\bibnamefont{{Mandelbaum}}}, \bibinfo{author}{\bibfnamefont{M.}~\bibnamefont{{Takada}}}, \bibinfo{author}{\bibfnamefont{S.}~\bibnamefont{{More}}}, \bibinfo{author}{\bibfnamefont{M.~A.} \bibnamefont{{Strauss}}}, \bibnamefont{et~al.}, \bibinfo{journal}{\prd} \textbf{\bibinfo{volume}{108}}, \bibinfo{eid}{123518} (\bibinfo{year}{2023}), \eprint{2304.00702}.

\bibitem[{\citenamefont{{Nguyen} et~al.}(2023)\citenamefont{{Nguyen}, {Huterer}, and {Wen}}}]{Nguyen2023EvidenceFor}
\bibinfo{author}{\bibfnamefont{N.-M.} \bibnamefont{{Nguyen}}}, \bibinfo{author}{\bibfnamefont{D.}~\bibnamefont{{Huterer}}}, \bibnamefont{and} \bibinfo{author}{\bibfnamefont{Y.}~\bibnamefont{{Wen}}}, \bibinfo{journal}{\prl} \textbf{\bibinfo{volume}{131}}, \bibinfo{eid}{111001} (\bibinfo{year}{2023}), \eprint{2302.01331}.

\bibitem[{\citenamefont{{Rogers} and {Poulin}}(2025)}]{2025PhRvR...7a2018R}
\bibinfo{author}{\bibfnamefont{K.~K.} \bibnamefont{{Rogers}}} \bibnamefont{and} \bibinfo{author}{\bibfnamefont{V.}~\bibnamefont{{Poulin}}}, \bibinfo{journal}{Physical Review Research} \textbf{\bibinfo{volume}{7}}, \bibinfo{eid}{L012018} (\bibinfo{year}{2025}), \eprint{2311.16377}.

\bibitem[{\citenamefont{{Fernandez} et~al.}(2024)\citenamefont{{Fernandez}, {Bird}, and {Ho}}}]{Fernandez2024CosmologicalConstraints}
\bibinfo{author}{\bibfnamefont{M.~A.} \bibnamefont{{Fernandez}}}, \bibinfo{author}{\bibfnamefont{S.}~\bibnamefont{{Bird}}}, \bibnamefont{and} \bibinfo{author}{\bibfnamefont{M.-F.} \bibnamefont{{Ho}}}, \bibinfo{journal}{\jcap} \textbf{\bibinfo{volume}{2024}}, \bibinfo{eid}{029} (\bibinfo{year}{2024}), \eprint{2309.03943}.

\bibitem[{\citenamefont{{Planck Collaboration} et~al.}(2020{\natexlab{b}})\citenamefont{{Planck Collaboration}, {Aghanim}, {Akrami}, {Ashdown}, {Aumont}, {Baccigalupi}, {Ballardini}, {Banday}, {Barreiro}, {Bartolo} et~al.}}]{Planck2018Lensing}
\bibinfo{author}{\bibnamefont{{Planck Collaboration}}}, \bibinfo{author}{\bibfnamefont{N.}~\bibnamefont{{Aghanim}}}, \bibinfo{author}{\bibfnamefont{Y.}~\bibnamefont{{Akrami}}}, \bibinfo{author}{\bibfnamefont{M.}~\bibnamefont{{Ashdown}}}, \bibinfo{author}{\bibfnamefont{J.}~\bibnamefont{{Aumont}}}, \bibinfo{author}{\bibfnamefont{C.}~\bibnamefont{{Baccigalupi}}}, \bibinfo{author}{\bibfnamefont{M.}~\bibnamefont{{Ballardini}}}, \bibinfo{author}{\bibfnamefont{A.~J.} \bibnamefont{{Banday}}}, \bibinfo{author}{\bibfnamefont{R.~B.} \bibnamefont{{Barreiro}}}, \bibinfo{author}{\bibfnamefont{N.}~\bibnamefont{{Bartolo}}}, \bibnamefont{et~al.}, \bibinfo{journal}{\aap} \textbf{\bibinfo{volume}{641}}, \bibinfo{eid}{A8} (\bibinfo{year}{2020}{\natexlab{b}}), \eprint{1807.06210}.

\bibitem[{\citenamefont{{Madhavacheril} et~al.}(2024)\citenamefont{{Madhavacheril}, {Qu}, {Sherwin}, {MacCrann}, {Li}, {Abril-Cabezas}, {Ade}, {Aiola}, {Alford}, {Amiri} et~al.}}]{Madhavacheril_2024}
\bibinfo{author}{\bibfnamefont{M.~S.} \bibnamefont{{Madhavacheril}}}, \bibinfo{author}{\bibfnamefont{F.~J.} \bibnamefont{{Qu}}}, \bibinfo{author}{\bibfnamefont{B.~D.} \bibnamefont{{Sherwin}}}, \bibinfo{author}{\bibfnamefont{N.}~\bibnamefont{{MacCrann}}}, \bibinfo{author}{\bibfnamefont{Y.}~\bibnamefont{{Li}}}, \bibinfo{author}{\bibfnamefont{I.}~\bibnamefont{{Abril-Cabezas}}}, \bibinfo{author}{\bibfnamefont{P.~A.~R.} \bibnamefont{{Ade}}}, \bibinfo{author}{\bibfnamefont{S.}~\bibnamefont{{Aiola}}}, \bibinfo{author}{\bibfnamefont{T.}~\bibnamefont{{Alford}}}, \bibinfo{author}{\bibfnamefont{M.}~\bibnamefont{{Amiri}}}, \bibnamefont{et~al.}, \bibinfo{journal}{\apj} \textbf{\bibinfo{volume}{962}}, \bibinfo{eid}{113} (\bibinfo{year}{2024}), \eprint{2304.05203}.

\bibitem[{\citenamefont{{Qu} et~al.}(2024)\citenamefont{{Qu}, {Sherwin}, {Madhavacheril}, {Han}, {Crowley}, {Abril-Cabezas}, {Ade}, {Aiola}, {Alford}, {Amiri} et~al.}}]{Qu_2024}
\bibinfo{author}{\bibfnamefont{F.~J.} \bibnamefont{{Qu}}}, \bibinfo{author}{\bibfnamefont{B.~D.} \bibnamefont{{Sherwin}}}, \bibinfo{author}{\bibfnamefont{M.~S.} \bibnamefont{{Madhavacheril}}}, \bibinfo{author}{\bibfnamefont{D.}~\bibnamefont{{Han}}}, \bibinfo{author}{\bibfnamefont{K.~T.} \bibnamefont{{Crowley}}}, \bibinfo{author}{\bibfnamefont{I.}~\bibnamefont{{Abril-Cabezas}}}, \bibinfo{author}{\bibfnamefont{P.~A.~R.} \bibnamefont{{Ade}}}, \bibinfo{author}{\bibfnamefont{S.}~\bibnamefont{{Aiola}}}, \bibinfo{author}{\bibfnamefont{T.}~\bibnamefont{{Alford}}}, \bibinfo{author}{\bibfnamefont{M.}~\bibnamefont{{Amiri}}}, \bibnamefont{et~al.}, \bibinfo{journal}{\apj} \textbf{\bibinfo{volume}{962}}, \bibinfo{eid}{112} (\bibinfo{year}{2024}), \eprint{2304.05202}.

\bibitem[{\citenamefont{{Bianchini} et~al.}(2020)\citenamefont{{Bianchini}, {Wu}, {Ade}, {Anderson}, {Austermann}, {Avva}, {Beall}, {Bender}, {Benson}, {Bleem} et~al.}}]{Bianchini2020SPTLensing}
\bibinfo{author}{\bibfnamefont{F.}~\bibnamefont{{Bianchini}}}, \bibinfo{author}{\bibfnamefont{W.~L.~K.} \bibnamefont{{Wu}}}, \bibinfo{author}{\bibfnamefont{P.~A.~R.} \bibnamefont{{Ade}}}, \bibinfo{author}{\bibfnamefont{A.~J.} \bibnamefont{{Anderson}}}, \bibinfo{author}{\bibfnamefont{J.~E.} \bibnamefont{{Austermann}}}, \bibinfo{author}{\bibfnamefont{J.~S.} \bibnamefont{{Avva}}}, \bibinfo{author}{\bibfnamefont{J.~A.} \bibnamefont{{Beall}}}, \bibinfo{author}{\bibfnamefont{A.~N.} \bibnamefont{{Bender}}}, \bibinfo{author}{\bibfnamefont{B.~A.} \bibnamefont{{Benson}}}, \bibinfo{author}{\bibfnamefont{L.~E.} \bibnamefont{{Bleem}}}, \bibnamefont{et~al.}, \bibinfo{journal}{\apj} \textbf{\bibinfo{volume}{888}}, \bibinfo{eid}{119} (\bibinfo{year}{2020}), \eprint{1910.07157}.

\bibitem[{\citenamefont{{Simpson}}(2010)}]{Simpson2010ScatteringOf}
\bibinfo{author}{\bibfnamefont{F.}~\bibnamefont{{Simpson}}}, \bibinfo{journal}{\prd} \textbf{\bibinfo{volume}{82}}, \bibinfo{eid}{083505} (\bibinfo{year}{2010}), \eprint{1007.1034}.

\bibitem[{\citenamefont{{Poulin} et~al.}(2023)\citenamefont{{Poulin}, {Bernal}, {Kovetz}, and {Kamionkowski}}}]{Poulin2023Sigma8Is}
\bibinfo{author}{\bibfnamefont{V.}~\bibnamefont{{Poulin}}}, \bibinfo{author}{\bibfnamefont{J.~L.} \bibnamefont{{Bernal}}}, \bibinfo{author}{\bibfnamefont{E.~D.} \bibnamefont{{Kovetz}}}, \bibnamefont{and} \bibinfo{author}{\bibfnamefont{M.}~\bibnamefont{{Kamionkowski}}}, \bibinfo{journal}{\prd} \textbf{\bibinfo{volume}{107}}, \bibinfo{eid}{123538} (\bibinfo{year}{2023}), \eprint{2209.06217}.

\bibitem[{\citenamefont{{Lagu{\"e}} et~al.}(2024)\citenamefont{{Lagu{\"e}}, {McCarthy}, {Madhavacheril}, {Hill}, and {Qu}}}]{Lague2024ConstraintsOn}
\bibinfo{author}{\bibfnamefont{A.}~\bibnamefont{{Lagu{\"e}}}}, \bibinfo{author}{\bibfnamefont{F.}~\bibnamefont{{McCarthy}}}, \bibinfo{author}{\bibfnamefont{M.}~\bibnamefont{{Madhavacheril}}}, \bibinfo{author}{\bibfnamefont{J.~C.} \bibnamefont{{Hill}}}, \bibnamefont{and} \bibinfo{author}{\bibfnamefont{F.~J.} \bibnamefont{{Qu}}}, \bibinfo{journal}{\prd} \textbf{\bibinfo{volume}{110}}, \bibinfo{eid}{023536} (\bibinfo{year}{2024}), \eprint{2402.08149}.

\bibitem[{\citenamefont{{Mazoun} et~al.}(2024)\citenamefont{{Mazoun}, {Bocquet}, {Garny}, {Mohr}, {Rubira}, and {Vogt}}}]{Mazoun2024ProbingInteracting}
\bibinfo{author}{\bibfnamefont{A.}~\bibnamefont{{Mazoun}}}, \bibinfo{author}{\bibfnamefont{S.}~\bibnamefont{{Bocquet}}}, \bibinfo{author}{\bibfnamefont{M.}~\bibnamefont{{Garny}}}, \bibinfo{author}{\bibfnamefont{J.~J.} \bibnamefont{{Mohr}}}, \bibinfo{author}{\bibfnamefont{H.}~\bibnamefont{{Rubira}}}, \bibnamefont{and} \bibinfo{author}{\bibfnamefont{S.~M.~L.} \bibnamefont{{Vogt}}}, \bibinfo{journal}{\prd} \textbf{\bibinfo{volume}{109}}, \bibinfo{eid}{063536} (\bibinfo{year}{2024}), \eprint{2312.17622}.

\bibitem[{\citenamefont{{Cyr-Racine} et~al.}(2016)\citenamefont{{Cyr-Racine}, {Sigurdson}, {Zavala}, {Bringmann}, {Vogelsberger}, and {Pfrommer}}}]{Cyr_Racine_2016}
\bibinfo{author}{\bibfnamefont{F.-Y.} \bibnamefont{{Cyr-Racine}}}, \bibinfo{author}{\bibfnamefont{K.}~\bibnamefont{{Sigurdson}}}, \bibinfo{author}{\bibfnamefont{J.}~\bibnamefont{{Zavala}}}, \bibinfo{author}{\bibfnamefont{T.}~\bibnamefont{{Bringmann}}}, \bibinfo{author}{\bibfnamefont{M.}~\bibnamefont{{Vogelsberger}}}, \bibnamefont{and} \bibinfo{author}{\bibfnamefont{C.}~\bibnamefont{{Pfrommer}}}, \bibinfo{journal}{\prd} \textbf{\bibinfo{volume}{93}}, \bibinfo{eid}{123527} (\bibinfo{year}{2016}), \eprint{1512.05344}.

\bibitem[{\citenamefont{{Markovic} et~al.}(2011)\citenamefont{{Markovic}, {Bridle}, {Slosar}, and {Weller}}}]{Markovic_2011}
\bibinfo{author}{\bibfnamefont{K.}~\bibnamefont{{Markovic}}}, \bibinfo{author}{\bibfnamefont{S.}~\bibnamefont{{Bridle}}}, \bibinfo{author}{\bibfnamefont{A.}~\bibnamefont{{Slosar}}}, \bibnamefont{and} \bibinfo{author}{\bibfnamefont{J.}~\bibnamefont{{Weller}}}, \bibinfo{journal}{\jcap} \textbf{\bibinfo{volume}{2011}}, \bibinfo{eid}{022} (\bibinfo{year}{2011}), \eprint{1009.0218}.

\bibitem[{\citenamefont{{Tulin} and {Yu}}(2018)}]{Tulin_2018}
\bibinfo{author}{\bibfnamefont{S.}~\bibnamefont{{Tulin}}} \bibnamefont{and} \bibinfo{author}{\bibfnamefont{H.-B.} \bibnamefont{{Yu}}}, \bibinfo{journal}{\physrep} \textbf{\bibinfo{volume}{730}}, \bibinfo{pages}{1} (\bibinfo{year}{2018}), \eprint{1705.02358}.

\bibitem[{\citenamefont{{Tsujikawa}}(2007)}]{Tsujikawa2007MatterDensity}
\bibinfo{author}{\bibfnamefont{S.}~\bibnamefont{{Tsujikawa}}}, \bibinfo{journal}{\prd} \textbf{\bibinfo{volume}{76}}, \bibinfo{eid}{023514} (\bibinfo{year}{2007}), \eprint{0705.1032}.

\bibitem[{\citenamefont{{Shah} et~al.}(2025)\citenamefont{{Shah}, {Mukherjee}, and {Pal}}}]{Shah2024ReconcilingS8}
\bibinfo{author}{\bibfnamefont{R.}~\bibnamefont{{Shah}}}, \bibinfo{author}{\bibfnamefont{P.}~\bibnamefont{{Mukherjee}}}, \bibnamefont{and} \bibinfo{author}{\bibfnamefont{S.}~\bibnamefont{{Pal}}}, \bibinfo{journal}{\mnras} \textbf{\bibinfo{volume}{536}}, \bibinfo{pages}{2404} (\bibinfo{year}{2025}), \eprint{2404.06396}.

\bibitem[{\citenamefont{{Wang} and {Steinhardt}}(1998)}]{Wang1998ClusterAbundance}
\bibinfo{author}{\bibfnamefont{L.}~\bibnamefont{{Wang}}} \bibnamefont{and} \bibinfo{author}{\bibfnamefont{P.~J.} \bibnamefont{{Steinhardt}}}, \bibinfo{journal}{\apj} \textbf{\bibinfo{volume}{508}}, \bibinfo{pages}{483} (\bibinfo{year}{1998}), \eprint{astro-ph/9804015}.

\bibitem[{\citenamefont{{Linder}}(2005)}]{Linder2005CosmicGrowth}
\bibinfo{author}{\bibfnamefont{E.~V.} \bibnamefont{{Linder}}}, \bibinfo{journal}{\prd} \textbf{\bibinfo{volume}{72}}, \bibinfo{eid}{043529} (\bibinfo{year}{2005}), \eprint{astro-ph/0507263}.

\bibitem[{\citenamefont{{Amon} and {Efstathiou}}(2022)}]{amon_2022}
\bibinfo{author}{\bibfnamefont{A.}~\bibnamefont{{Amon}}} \bibnamefont{and} \bibinfo{author}{\bibfnamefont{G.}~\bibnamefont{{Efstathiou}}}, \bibinfo{journal}{\mnras} \textbf{\bibinfo{volume}{516}}, \bibinfo{pages}{5355} (\bibinfo{year}{2022}), \eprint{2206.11794}.

\bibitem[{\citenamefont{Schaan et~al.}(2021)\citenamefont{Schaan, Ferraro, Amodeo, Battaglia, Aiola, Austermann, Beall, Bean, Becker, Bond et~al.}}]{Schaan_2021}
\bibinfo{author}{\bibfnamefont{E.}~\bibnamefont{Schaan}}, \bibinfo{author}{\bibfnamefont{S.}~\bibnamefont{Ferraro}}, \bibinfo{author}{\bibfnamefont{S.}~\bibnamefont{Amodeo}}, \bibinfo{author}{\bibfnamefont{N.}~\bibnamefont{Battaglia}}, \bibinfo{author}{\bibfnamefont{S.}~\bibnamefont{Aiola}}, \bibinfo{author}{\bibfnamefont{J.~E.} \bibnamefont{Austermann}}, \bibinfo{author}{\bibfnamefont{J.~A.} \bibnamefont{Beall}}, \bibinfo{author}{\bibfnamefont{R.}~\bibnamefont{Bean}}, \bibinfo{author}{\bibfnamefont{D.~T.} \bibnamefont{Becker}}, \bibinfo{author}{\bibfnamefont{R.~J.} \bibnamefont{Bond}}, \bibnamefont{et~al.}, \bibinfo{journal}{Physical Review D} \textbf{\bibinfo{volume}{103}} (\bibinfo{year}{2021}), ISSN \bibinfo{issn}{2470-0029}, \urlprefix\url{http://dx.doi.org/10.1103/PhysRevD.103.063513}.

\bibitem[{\citenamefont{{Amodeo} et~al.}(2021)\citenamefont{{Amodeo}, {Battaglia}, {Schaan}, {Ferraro}, {Moser}, {Aiola}, {Austermann}, {Beall}, {Bean}, {Becker} et~al.}}]{Amodeo_2021}
\bibinfo{author}{\bibfnamefont{S.}~\bibnamefont{{Amodeo}}}, \bibinfo{author}{\bibfnamefont{N.}~\bibnamefont{{Battaglia}}}, \bibinfo{author}{\bibfnamefont{E.}~\bibnamefont{{Schaan}}}, \bibinfo{author}{\bibfnamefont{S.}~\bibnamefont{{Ferraro}}}, \bibinfo{author}{\bibfnamefont{E.}~\bibnamefont{{Moser}}}, \bibinfo{author}{\bibfnamefont{S.}~\bibnamefont{{Aiola}}}, \bibinfo{author}{\bibfnamefont{J.~E.} \bibnamefont{{Austermann}}}, \bibinfo{author}{\bibfnamefont{J.~A.} \bibnamefont{{Beall}}}, \bibinfo{author}{\bibfnamefont{R.}~\bibnamefont{{Bean}}}, \bibinfo{author}{\bibfnamefont{D.~T.} \bibnamefont{{Becker}}}, \bibnamefont{et~al.}, \bibinfo{journal}{\prd} \textbf{\bibinfo{volume}{103}}, \bibinfo{eid}{063514} (\bibinfo{year}{2021}), \eprint{2009.05558}.

\bibitem[{\citenamefont{{Hadzhiyska} et~al.}(2024{\natexlab{a}})\citenamefont{{Hadzhiyska}, {Ferraro}, {Ried Guachalla}, {Schaan}, {Aguilar}, {Battaglia}, {Bond}, {Brooks}, {Calabrese}, {Choi} et~al.}}]{boryana_ksz}
\bibinfo{author}{\bibfnamefont{B.}~\bibnamefont{{Hadzhiyska}}}, \bibinfo{author}{\bibfnamefont{S.}~\bibnamefont{{Ferraro}}}, \bibinfo{author}{\bibfnamefont{B.}~\bibnamefont{{Ried Guachalla}}}, \bibinfo{author}{\bibfnamefont{E.}~\bibnamefont{{Schaan}}}, \bibinfo{author}{\bibfnamefont{J.}~\bibnamefont{{Aguilar}}}, \bibinfo{author}{\bibfnamefont{N.}~\bibnamefont{{Battaglia}}}, \bibinfo{author}{\bibfnamefont{J.~R.} \bibnamefont{{Bond}}}, \bibinfo{author}{\bibfnamefont{D.}~\bibnamefont{{Brooks}}}, \bibinfo{author}{\bibfnamefont{E.}~\bibnamefont{{Calabrese}}}, \bibinfo{author}{\bibfnamefont{S.~K.} \bibnamefont{{Choi}}}, \bibnamefont{et~al.}, \bibinfo{journal}{arXiv e-prints} \bibinfo{eid}{arXiv:2407.07152} (\bibinfo{year}{2024}{\natexlab{a}}), \eprint{2407.07152}.

\bibitem[{\citenamefont{{Schneider} et~al.}(2022)\citenamefont{{Schneider}, {Giri}, {Amodeo}, and {Refregier}}}]{Schneider2022ConstrainingBaryons}
\bibinfo{author}{\bibfnamefont{A.}~\bibnamefont{{Schneider}}}, \bibinfo{author}{\bibfnamefont{S.~K.} \bibnamefont{{Giri}}}, \bibinfo{author}{\bibfnamefont{S.}~\bibnamefont{{Amodeo}}}, \bibnamefont{and} \bibinfo{author}{\bibfnamefont{A.}~\bibnamefont{{Refregier}}}, \bibinfo{journal}{\mnras} \textbf{\bibinfo{volume}{514}}, \bibinfo{pages}{3802} (\bibinfo{year}{2022}), \eprint{2110.02228}.

\bibitem[{\citenamefont{{Tegmark} and {Zaldarriaga}}(2002)}]{Tegmark2002SeparatingThe}
\bibinfo{author}{\bibfnamefont{M.}~\bibnamefont{{Tegmark}}} \bibnamefont{and} \bibinfo{author}{\bibfnamefont{M.}~\bibnamefont{{Zaldarriaga}}}, \bibinfo{journal}{\prd} \textbf{\bibinfo{volume}{66}}, \bibinfo{eid}{103508} (\bibinfo{year}{2002}), \eprint{astro-ph/0207047}.

\bibitem[{\citenamefont{{Chabanier} et~al.}(2019)\citenamefont{{Chabanier}, {Millea}, and {Palanque-Delabrouille}}}]{Chabanier2019MatterPower}
\bibinfo{author}{\bibfnamefont{S.}~\bibnamefont{{Chabanier}}}, \bibinfo{author}{\bibfnamefont{M.}~\bibnamefont{{Millea}}}, \bibnamefont{and} \bibinfo{author}{\bibfnamefont{N.}~\bibnamefont{{Palanque-Delabrouille}}}, \bibinfo{journal}{\mnras} \textbf{\bibinfo{volume}{489}}, \bibinfo{pages}{2247} (\bibinfo{year}{2019}), \eprint{1905.08103}.

\bibitem[{\citenamefont{{Doux} et~al.}(2022)\citenamefont{{Doux}, {Jain}, {Zeurcher}, {Lee}, {Fang}, {Rosenfeld}, {Amon}, {Camacho}, {Choi}, {Secco} et~al.}}]{Doux2022DES}
\bibinfo{author}{\bibfnamefont{C.}~\bibnamefont{{Doux}}}, \bibinfo{author}{\bibfnamefont{B.}~\bibnamefont{{Jain}}}, \bibinfo{author}{\bibfnamefont{D.}~\bibnamefont{{Zeurcher}}}, \bibinfo{author}{\bibfnamefont{J.}~\bibnamefont{{Lee}}}, \bibinfo{author}{\bibfnamefont{X.}~\bibnamefont{{Fang}}}, \bibinfo{author}{\bibfnamefont{R.}~\bibnamefont{{Rosenfeld}}}, \bibinfo{author}{\bibfnamefont{A.}~\bibnamefont{{Amon}}}, \bibinfo{author}{\bibfnamefont{H.}~\bibnamefont{{Camacho}}}, \bibinfo{author}{\bibfnamefont{A.}~\bibnamefont{{Choi}}}, \bibinfo{author}{\bibfnamefont{L.~F.} \bibnamefont{{Secco}}}, \bibnamefont{et~al.}, \bibinfo{journal}{\mnras} \textbf{\bibinfo{volume}{515}}, \bibinfo{pages}{1942} (\bibinfo{year}{2022}), \eprint{2203.07128}.

\bibitem[{\citenamefont{{Broxterman} and {Kuijken}}(2024)}]{2409.13404}
\bibinfo{author}{\bibfnamefont{J.~C.} \bibnamefont{{Broxterman}}} \bibnamefont{and} \bibinfo{author}{\bibfnamefont{K.}~\bibnamefont{{Kuijken}}}, \bibinfo{journal}{arXiv e-prints} \bibinfo{eid}{arXiv:2409.13404} (\bibinfo{year}{2024}), \eprint{2409.13404}.

\bibitem[{\citenamefont{{Preston} et~al.}(2023)\citenamefont{{Preston}, {Amon}, and {Efstathiou}}}]{preston_2023}
\bibinfo{author}{\bibfnamefont{C.}~\bibnamefont{{Preston}}}, \bibinfo{author}{\bibfnamefont{A.}~\bibnamefont{{Amon}}}, \bibnamefont{and} \bibinfo{author}{\bibfnamefont{G.}~\bibnamefont{{Efstathiou}}}, \bibinfo{journal}{\mnras} \textbf{\bibinfo{volume}{525}}, \bibinfo{pages}{5554} (\bibinfo{year}{2023}), \eprint{2305.09827}.

\bibitem[{\citenamefont{{Preston} et~al.}(2024)\citenamefont{{Preston}, {Amon}, and {Efstathiou}}}]{preston_2024}
\bibinfo{author}{\bibfnamefont{C.}~\bibnamefont{{Preston}}}, \bibinfo{author}{\bibfnamefont{A.}~\bibnamefont{{Amon}}}, \bibnamefont{and} \bibinfo{author}{\bibfnamefont{G.}~\bibnamefont{{Efstathiou}}}, \bibinfo{journal}{\mnras} \textbf{\bibinfo{volume}{533}}, \bibinfo{pages}{621} (\bibinfo{year}{2024}), \eprint{2404.18240}.

\bibitem[{\citenamefont{{Secco} et~al.}(2022)\citenamefont{{Secco}, {Samuroff}, {Krause}, {Jain}, {Blazek}, {Raveri}, {Campos}, {Amon}, {Chen}, {Doux} et~al.}}]{Secco2022CosmicShear}
\bibinfo{author}{\bibfnamefont{L.~F.} \bibnamefont{{Secco}}}, \bibinfo{author}{\bibfnamefont{S.}~\bibnamefont{{Samuroff}}}, \bibinfo{author}{\bibfnamefont{E.}~\bibnamefont{{Krause}}}, \bibinfo{author}{\bibfnamefont{B.}~\bibnamefont{{Jain}}}, \bibinfo{author}{\bibfnamefont{J.}~\bibnamefont{{Blazek}}}, \bibinfo{author}{\bibfnamefont{M.}~\bibnamefont{{Raveri}}}, \bibinfo{author}{\bibfnamefont{A.}~\bibnamefont{{Campos}}}, \bibinfo{author}{\bibfnamefont{A.}~\bibnamefont{{Amon}}}, \bibinfo{author}{\bibfnamefont{A.}~\bibnamefont{{Chen}}}, \bibinfo{author}{\bibfnamefont{C.}~\bibnamefont{{Doux}}}, \bibnamefont{et~al.}, \bibinfo{journal}{\prd} \textbf{\bibinfo{volume}{105}}, \bibinfo{eid}{023515} (\bibinfo{year}{2022}), \eprint{2105.13544}.

\bibitem[{\citenamefont{{Krause} et~al.}(2021)\citenamefont{{Krause}, {Fang}, {Pandey}, {Secco}, {Alves}, {Huang}, {Blazek}, {Prat}, {Zuntz}, {Eifler} et~al.}}]{krause2021darkenergysurveyyear}
\bibinfo{author}{\bibfnamefont{E.}~\bibnamefont{{Krause}}}, \bibinfo{author}{\bibfnamefont{X.}~\bibnamefont{{Fang}}}, \bibinfo{author}{\bibfnamefont{S.}~\bibnamefont{{Pandey}}}, \bibinfo{author}{\bibfnamefont{L.~F.} \bibnamefont{{Secco}}}, \bibinfo{author}{\bibfnamefont{O.}~\bibnamefont{{Alves}}}, \bibinfo{author}{\bibfnamefont{H.}~\bibnamefont{{Huang}}}, \bibinfo{author}{\bibfnamefont{J.}~\bibnamefont{{Blazek}}}, \bibinfo{author}{\bibfnamefont{J.}~\bibnamefont{{Prat}}}, \bibinfo{author}{\bibfnamefont{J.}~\bibnamefont{{Zuntz}}}, \bibinfo{author}{\bibfnamefont{T.~F.} \bibnamefont{{Eifler}}}, \bibnamefont{et~al.}, \bibinfo{journal}{arXiv e-prints} \bibinfo{eid}{arXiv:2105.13548} (\bibinfo{year}{2021}), \eprint{2105.13548}.

\bibitem[{\citenamefont{{Ye} et~al.}(2024)\citenamefont{{Ye}, {Jiang}, and {Silvestri}}}]{ye2024}
\bibinfo{author}{\bibfnamefont{G.}~\bibnamefont{{Ye}}}, \bibinfo{author}{\bibfnamefont{J.-Q.} \bibnamefont{{Jiang}}}, \bibnamefont{and} \bibinfo{author}{\bibfnamefont{A.}~\bibnamefont{{Silvestri}}}, \bibinfo{journal}{arXiv e-prints} \bibinfo{eid}{arXiv:2411.07082} (\bibinfo{year}{2024}), \eprint{2411.07082}.

\bibitem[{\citenamefont{{Zhong} et~al.}(2023)\citenamefont{{Zhong}, {Saraivanov}, {Miranda}, {Xu}, {Eifler}, and {Krause}}}]{Zhong2023GrowthAnd}
\bibinfo{author}{\bibfnamefont{K.}~\bibnamefont{{Zhong}}}, \bibinfo{author}{\bibfnamefont{E.}~\bibnamefont{{Saraivanov}}}, \bibinfo{author}{\bibfnamefont{V.}~\bibnamefont{{Miranda}}}, \bibinfo{author}{\bibfnamefont{J.}~\bibnamefont{{Xu}}}, \bibinfo{author}{\bibfnamefont{T.}~\bibnamefont{{Eifler}}}, \bibnamefont{and} \bibinfo{author}{\bibfnamefont{E.}~\bibnamefont{{Krause}}}, \bibinfo{journal}{\prd} \textbf{\bibinfo{volume}{107}}, \bibinfo{eid}{123529} (\bibinfo{year}{2023}), \eprint{2301.03694}.

\bibitem[{\citenamefont{{Krause} and {Eifler}}(2017)}]{Krause2017cosmolike}
\bibinfo{author}{\bibfnamefont{E.}~\bibnamefont{{Krause}}} \bibnamefont{and} \bibinfo{author}{\bibfnamefont{T.}~\bibnamefont{{Eifler}}}, \bibinfo{journal}{\mnras} \textbf{\bibinfo{volume}{470}}, \bibinfo{pages}{2100} (\bibinfo{year}{2017}), \eprint{1601.05779}.

\bibitem[{\citenamefont{{Amon} et~al.}(2022)\citenamefont{{Amon}, {Gruen}, {Troxel}, {MacCrann}, {Dodelson}, {Choi}, {Doux}, {Secco}, {Samuroff}, {Krause} et~al.}}]{DESY3_cosmicshear_amon}
\bibinfo{author}{\bibfnamefont{A.}~\bibnamefont{{Amon}}}, \bibinfo{author}{\bibfnamefont{D.}~\bibnamefont{{Gruen}}}, \bibinfo{author}{\bibfnamefont{M.~A.} \bibnamefont{{Troxel}}}, \bibinfo{author}{\bibfnamefont{N.}~\bibnamefont{{MacCrann}}}, \bibinfo{author}{\bibfnamefont{S.}~\bibnamefont{{Dodelson}}}, \bibinfo{author}{\bibfnamefont{A.}~\bibnamefont{{Choi}}}, \bibinfo{author}{\bibfnamefont{C.}~\bibnamefont{{Doux}}}, \bibinfo{author}{\bibfnamefont{L.~F.} \bibnamefont{{Secco}}}, \bibinfo{author}{\bibfnamefont{S.}~\bibnamefont{{Samuroff}}}, \bibinfo{author}{\bibfnamefont{E.}~\bibnamefont{{Krause}}}, \bibnamefont{et~al.}, \bibinfo{journal}{\prd} \textbf{\bibinfo{volume}{105}}, \bibinfo{eid}{023514} (\bibinfo{year}{2022}), \eprint{2105.13543}.

\bibitem[{\citenamefont{{Lewis} et~al.}(2000)\citenamefont{{Lewis}, {Challinor}, and {Lasenby}}}]{CAMB}
\bibinfo{author}{\bibfnamefont{A.}~\bibnamefont{{Lewis}}}, \bibinfo{author}{\bibfnamefont{A.}~\bibnamefont{{Challinor}}}, \bibnamefont{and} \bibinfo{author}{\bibfnamefont{A.}~\bibnamefont{{Lasenby}}}, \bibinfo{journal}{\apj} \textbf{\bibinfo{volume}{538}}, \bibinfo{pages}{473} (\bibinfo{year}{2000}), \eprint{astro-ph/9911177}.

\bibitem[{\citenamefont{{Mead} et~al.}(2021)\citenamefont{{Mead}, {Brieden}, {Tr{\"o}ster}, and {Heymans}}}]{Mead_2020}
\bibinfo{author}{\bibfnamefont{A.~J.} \bibnamefont{{Mead}}}, \bibinfo{author}{\bibfnamefont{S.}~\bibnamefont{{Brieden}}}, \bibinfo{author}{\bibfnamefont{T.}~\bibnamefont{{Tr{\"o}ster}}}, \bibnamefont{and} \bibinfo{author}{\bibfnamefont{C.}~\bibnamefont{{Heymans}}}, \bibinfo{journal}{\mnras} \textbf{\bibinfo{volume}{502}}, \bibinfo{pages}{1401} (\bibinfo{year}{2021}), \eprint{2009.01858}.

\bibitem[{\citenamefont{{Blazek} et~al.}(2019)\citenamefont{{Blazek}, {MacCrann}, {Troxel}, and {Fang}}}]{Blazek_2019}
\bibinfo{author}{\bibfnamefont{J.~A.} \bibnamefont{{Blazek}}}, \bibinfo{author}{\bibfnamefont{N.}~\bibnamefont{{MacCrann}}}, \bibinfo{author}{\bibfnamefont{M.~A.} \bibnamefont{{Troxel}}}, \bibnamefont{and} \bibinfo{author}{\bibfnamefont{X.}~\bibnamefont{{Fang}}}, \bibinfo{journal}{\prd} \textbf{\bibinfo{volume}{100}}, \bibinfo{eid}{103506} (\bibinfo{year}{2019}), \eprint{1708.09247}.

\bibitem[{\citenamefont{{Salcido} et~al.}(2023)\citenamefont{{Salcido}, {McCarthy}, {Kwan}, {Upadhye}, and {Font}}}]{Salcido_2023}
\bibinfo{author}{\bibfnamefont{J.}~\bibnamefont{{Salcido}}}, \bibinfo{author}{\bibfnamefont{I.~G.} \bibnamefont{{McCarthy}}}, \bibinfo{author}{\bibfnamefont{J.}~\bibnamefont{{Kwan}}}, \bibinfo{author}{\bibfnamefont{A.}~\bibnamefont{{Upadhye}}}, \bibnamefont{and} \bibinfo{author}{\bibfnamefont{A.~S.} \bibnamefont{{Font}}}, \bibinfo{journal}{\mnras} \textbf{\bibinfo{volume}{523}}, \bibinfo{pages}{2247} (\bibinfo{year}{2023}), \eprint{2305.09710}.

\bibitem[{\citenamefont{{Lagu{\"e}} et~al.}(2022)\citenamefont{{Lagu{\"e}}, {Bond}, {Hlo{\v{z}}ek}, {Rogers}, {Marsh}, and {Grin}}}]{Lague2022ConstrainingUltralight}
\bibinfo{author}{\bibfnamefont{A.}~\bibnamefont{{Lagu{\"e}}}}, \bibinfo{author}{\bibfnamefont{J.~R.} \bibnamefont{{Bond}}}, \bibinfo{author}{\bibfnamefont{R.}~\bibnamefont{{Hlo{\v{z}}ek}}}, \bibinfo{author}{\bibfnamefont{K.~K.} \bibnamefont{{Rogers}}}, \bibinfo{author}{\bibfnamefont{D.~J.~E.} \bibnamefont{{Marsh}}}, \bibnamefont{and} \bibinfo{author}{\bibfnamefont{D.}~\bibnamefont{{Grin}}}, \bibinfo{journal}{\jcap} \textbf{\bibinfo{volume}{2022}}, \bibinfo{eid}{049} (\bibinfo{year}{2022}), \eprint{2104.07802}.

\bibitem[{\citenamefont{{Rogers} et~al.}(2023)\citenamefont{{Rogers}, {Hlo{\v{z}}ek}, {Lagu{\"e}}, {Ivanov}, {Philcox}, {Cabass}, {Akitsu}, and {Marsh}}}]{Rogers2023UltralightAxions}
\bibinfo{author}{\bibfnamefont{K.~K.} \bibnamefont{{Rogers}}}, \bibinfo{author}{\bibfnamefont{R.}~\bibnamefont{{Hlo{\v{z}}ek}}}, \bibinfo{author}{\bibfnamefont{A.}~\bibnamefont{{Lagu{\"e}}}}, \bibinfo{author}{\bibfnamefont{M.~M.} \bibnamefont{{Ivanov}}}, \bibinfo{author}{\bibfnamefont{O.~H.~E.} \bibnamefont{{Philcox}}}, \bibinfo{author}{\bibfnamefont{G.}~\bibnamefont{{Cabass}}}, \bibinfo{author}{\bibfnamefont{K.}~\bibnamefont{{Akitsu}}}, \bibnamefont{and} \bibinfo{author}{\bibfnamefont{D.~J.~E.} \bibnamefont{{Marsh}}}, \bibinfo{journal}{\jcap} \textbf{\bibinfo{volume}{2023}}, \bibinfo{eid}{023} (\bibinfo{year}{2023}), \eprint{2301.08361}.

\bibitem[{\citenamefont{{Vogt} et~al.}(2023)\citenamefont{{Vogt}, {Marsh}, and {Lagu{\"e}}}}]{Vogt2023ImprovedMixed}
\bibinfo{author}{\bibfnamefont{S.~M.~L.} \bibnamefont{{Vogt}}}, \bibinfo{author}{\bibfnamefont{D.~J.~E.} \bibnamefont{{Marsh}}}, \bibnamefont{and} \bibinfo{author}{\bibfnamefont{A.}~\bibnamefont{{Lagu{\"e}}}}, \bibinfo{journal}{\prd} \textbf{\bibinfo{volume}{107}}, \bibinfo{eid}{063526} (\bibinfo{year}{2023}), \eprint{2209.13445}.

\bibitem[{\citenamefont{{Hlozek} et~al.}(2015)\citenamefont{{Hlozek}, {Grin}, {Marsh}, and {Ferreira}}}]{Hlozek2015ASearch}
\bibinfo{author}{\bibfnamefont{R.}~\bibnamefont{{Hlozek}}}, \bibinfo{author}{\bibfnamefont{D.}~\bibnamefont{{Grin}}}, \bibinfo{author}{\bibfnamefont{D.~J.~E.} \bibnamefont{{Marsh}}}, \bibnamefont{and} \bibinfo{author}{\bibfnamefont{P.~G.} \bibnamefont{{Ferreira}}}, \bibinfo{journal}{\prd} \textbf{\bibinfo{volume}{91}}, \bibinfo{eid}{103512} (\bibinfo{year}{2015}), \eprint{1410.2896}.

\bibitem[{\citenamefont{{Gatti} et~al.}(2022)\citenamefont{{Gatti}, {Pandey}, {Baxter}, {Hill}, {Moser}, {Raveri}, {Fang}, {DeRose}, {Giannini}, {Doux} et~al.}}]{2108.01600}
\bibinfo{author}{\bibfnamefont{M.}~\bibnamefont{{Gatti}}}, \bibinfo{author}{\bibfnamefont{S.}~\bibnamefont{{Pandey}}}, \bibinfo{author}{\bibfnamefont{E.}~\bibnamefont{{Baxter}}}, \bibinfo{author}{\bibfnamefont{J.~C.} \bibnamefont{{Hill}}}, \bibinfo{author}{\bibfnamefont{E.}~\bibnamefont{{Moser}}}, \bibinfo{author}{\bibfnamefont{M.}~\bibnamefont{{Raveri}}}, \bibinfo{author}{\bibfnamefont{X.}~\bibnamefont{{Fang}}}, \bibinfo{author}{\bibfnamefont{J.}~\bibnamefont{{DeRose}}}, \bibinfo{author}{\bibfnamefont{G.}~\bibnamefont{{Giannini}}}, \bibinfo{author}{\bibfnamefont{C.}~\bibnamefont{{Doux}}}, \bibnamefont{et~al.}, \bibinfo{journal}{\prd} \textbf{\bibinfo{volume}{105}}, \bibinfo{eid}{123525} (\bibinfo{year}{2022}), \eprint{2108.01600}.

\bibitem[{\citenamefont{{Pandey} et~al.}(2022)\citenamefont{{Pandey}, {Gatti}, {Baxter}, {Hill}, {Fang}, {Doux}, {Giannini}, {Raveri}, {DeRose}, {Huang} et~al.}}]{2108.01601}
\bibinfo{author}{\bibfnamefont{S.}~\bibnamefont{{Pandey}}}, \bibinfo{author}{\bibfnamefont{M.}~\bibnamefont{{Gatti}}}, \bibinfo{author}{\bibfnamefont{E.}~\bibnamefont{{Baxter}}}, \bibinfo{author}{\bibfnamefont{J.~C.} \bibnamefont{{Hill}}}, \bibinfo{author}{\bibfnamefont{X.}~\bibnamefont{{Fang}}}, \bibinfo{author}{\bibfnamefont{C.}~\bibnamefont{{Doux}}}, \bibinfo{author}{\bibfnamefont{G.}~\bibnamefont{{Giannini}}}, \bibinfo{author}{\bibfnamefont{M.}~\bibnamefont{{Raveri}}}, \bibinfo{author}{\bibfnamefont{J.}~\bibnamefont{{DeRose}}}, \bibinfo{author}{\bibfnamefont{H.}~\bibnamefont{{Huang}}}, \bibnamefont{et~al.}, \bibinfo{journal}{\prd} \textbf{\bibinfo{volume}{105}}, \bibinfo{eid}{123526} (\bibinfo{year}{2022}), \eprint{2108.01601}.

\bibitem[{\citenamefont{{Tr{\"o}ster} et~al.}(2022)\citenamefont{{Tr{\"o}ster}, {Mead}, {Heymans}, {Yan}, {Alonso}, {Asgari}, {Bilicki}, {Dvornik}, {Hildebrandt}, {Joachimi} et~al.}}]{2109.04458}
\bibinfo{author}{\bibfnamefont{T.}~\bibnamefont{{Tr{\"o}ster}}}, \bibinfo{author}{\bibfnamefont{A.~J.} \bibnamefont{{Mead}}}, \bibinfo{author}{\bibfnamefont{C.}~\bibnamefont{{Heymans}}}, \bibinfo{author}{\bibfnamefont{Z.}~\bibnamefont{{Yan}}}, \bibinfo{author}{\bibfnamefont{D.}~\bibnamefont{{Alonso}}}, \bibinfo{author}{\bibfnamefont{M.}~\bibnamefont{{Asgari}}}, \bibinfo{author}{\bibfnamefont{M.}~\bibnamefont{{Bilicki}}}, \bibinfo{author}{\bibfnamefont{A.}~\bibnamefont{{Dvornik}}}, \bibinfo{author}{\bibfnamefont{H.}~\bibnamefont{{Hildebrandt}}}, \bibinfo{author}{\bibfnamefont{B.}~\bibnamefont{{Joachimi}}}, \bibnamefont{et~al.}, \bibinfo{journal}{\aap} \textbf{\bibinfo{volume}{660}}, \bibinfo{eid}{A27} (\bibinfo{year}{2022}), \eprint{2109.04458}.

\bibitem[{\citenamefont{{Pandey} et~al.}(2023)\citenamefont{{Pandey}, {Lehman}, {Baxter}, {Ni}, {Angl{\'e}s-Alc{\'a}zar}, {Genel}, {Villaescusa-Navarro}, {Delgado}, and {di Matteo}}}]{2301.02186}
\bibinfo{author}{\bibfnamefont{S.}~\bibnamefont{{Pandey}}}, \bibinfo{author}{\bibfnamefont{K.}~\bibnamefont{{Lehman}}}, \bibinfo{author}{\bibfnamefont{E.~J.} \bibnamefont{{Baxter}}}, \bibinfo{author}{\bibfnamefont{Y.}~\bibnamefont{{Ni}}}, \bibinfo{author}{\bibfnamefont{D.}~\bibnamefont{{Angl{\'e}s-Alc{\'a}zar}}}, \bibinfo{author}{\bibfnamefont{S.}~\bibnamefont{{Genel}}}, \bibinfo{author}{\bibfnamefont{F.}~\bibnamefont{{Villaescusa-Navarro}}}, \bibinfo{author}{\bibfnamefont{A.~M.} \bibnamefont{{Delgado}}}, \bibnamefont{and} \bibinfo{author}{\bibfnamefont{T.}~\bibnamefont{{di Matteo}}}, \bibinfo{journal}{\mnras} \textbf{\bibinfo{volume}{525}}, \bibinfo{pages}{1779} (\bibinfo{year}{2023}), \eprint{2301.02186}.

\bibitem[{\citenamefont{{Battaglia} et~al.}(2017)\citenamefont{{Battaglia}, {Ferraro}, {Schaan}, and {Spergel}}}]{1705.05881}
\bibinfo{author}{\bibfnamefont{N.}~\bibnamefont{{Battaglia}}}, \bibinfo{author}{\bibfnamefont{S.}~\bibnamefont{{Ferraro}}}, \bibinfo{author}{\bibfnamefont{E.}~\bibnamefont{{Schaan}}}, \bibnamefont{and} \bibinfo{author}{\bibfnamefont{D.~N.} \bibnamefont{{Spergel}}}, \bibinfo{journal}{\jcap} \textbf{\bibinfo{volume}{2017}}, \bibinfo{eid}{040} (\bibinfo{year}{2017}), \eprint{1705.05881}.

\bibitem[{\citenamefont{{Battaglia} et~al.}(2019)\citenamefont{{Battaglia}, {Hill}, {Amodeo}, {Bartlett}, {Basu}, {Erler}, {Ferraro}, {Hernquist}, {Madhavacheril}, {McQuinn} et~al.}}]{1903.04647}
\bibinfo{author}{\bibfnamefont{N.}~\bibnamefont{{Battaglia}}}, \bibinfo{author}{\bibfnamefont{J.~C.} \bibnamefont{{Hill}}}, \bibinfo{author}{\bibfnamefont{S.}~\bibnamefont{{Amodeo}}}, \bibinfo{author}{\bibfnamefont{J.~G.} \bibnamefont{{Bartlett}}}, \bibinfo{author}{\bibfnamefont{K.}~\bibnamefont{{Basu}}}, \bibinfo{author}{\bibfnamefont{J.}~\bibnamefont{{Erler}}}, \bibinfo{author}{\bibfnamefont{S.}~\bibnamefont{{Ferraro}}}, \bibinfo{author}{\bibfnamefont{L.}~\bibnamefont{{Hernquist}}}, \bibinfo{author}{\bibfnamefont{M.}~\bibnamefont{{Madhavacheril}}}, \bibinfo{author}{\bibfnamefont{M.}~\bibnamefont{{McQuinn}}}, \bibnamefont{et~al.}, \bibinfo{journal}{\baas} \textbf{\bibinfo{volume}{51}}, \bibinfo{eid}{297} (\bibinfo{year}{2019}), \eprint{1903.04647}.

\bibitem[{\citenamefont{{Schaan} et~al.}(2021)\citenamefont{{Schaan}, {Ferraro}, {Amodeo}, {Battaglia}, {Aiola}, {Austermann}, {Beall}, {Bean}, {Becker}, {Bond} et~al.}}]{2009.05557}
\bibinfo{author}{\bibfnamefont{E.}~\bibnamefont{{Schaan}}}, \bibinfo{author}{\bibfnamefont{S.}~\bibnamefont{{Ferraro}}}, \bibinfo{author}{\bibfnamefont{S.}~\bibnamefont{{Amodeo}}}, \bibinfo{author}{\bibfnamefont{N.}~\bibnamefont{{Battaglia}}}, \bibinfo{author}{\bibfnamefont{S.}~\bibnamefont{{Aiola}}}, \bibinfo{author}{\bibfnamefont{J.~E.} \bibnamefont{{Austermann}}}, \bibinfo{author}{\bibfnamefont{J.~A.} \bibnamefont{{Beall}}}, \bibinfo{author}{\bibfnamefont{R.}~\bibnamefont{{Bean}}}, \bibinfo{author}{\bibfnamefont{D.~T.} \bibnamefont{{Becker}}}, \bibinfo{author}{\bibfnamefont{R.~J.} \bibnamefont{{Bond}}}, \bibnamefont{et~al.}, \bibinfo{journal}{\prd} \textbf{\bibinfo{volume}{103}}, \bibinfo{eid}{063513} (\bibinfo{year}{2021}), \eprint{2009.05557}.

\bibitem[{\citenamefont{{Vavagiakis} et~al.}(2021)\citenamefont{{Vavagiakis}, {Gallardo}, {Calafut}, {Amodeo}, {Aiola}, {Austermann}, {Battaglia}, {Battistelli}, {Beall}, {Bean} et~al.}}]{2101.08373}
\bibinfo{author}{\bibfnamefont{E.~M.} \bibnamefont{{Vavagiakis}}}, \bibinfo{author}{\bibfnamefont{P.~A.} \bibnamefont{{Gallardo}}}, \bibinfo{author}{\bibfnamefont{V.}~\bibnamefont{{Calafut}}}, \bibinfo{author}{\bibfnamefont{S.}~\bibnamefont{{Amodeo}}}, \bibinfo{author}{\bibfnamefont{S.}~\bibnamefont{{Aiola}}}, \bibinfo{author}{\bibfnamefont{J.~E.} \bibnamefont{{Austermann}}}, \bibinfo{author}{\bibfnamefont{N.}~\bibnamefont{{Battaglia}}}, \bibinfo{author}{\bibfnamefont{E.~S.} \bibnamefont{{Battistelli}}}, \bibinfo{author}{\bibfnamefont{J.~A.} \bibnamefont{{Beall}}}, \bibinfo{author}{\bibfnamefont{R.}~\bibnamefont{{Bean}}}, \bibnamefont{et~al.}, \bibinfo{journal}{\prd} \textbf{\bibinfo{volume}{104}}, \bibinfo{eid}{043503} (\bibinfo{year}{2021}), \eprint{2101.08373}.

\bibitem[{\citenamefont{{Hadzhiyska} et~al.}(2024{\natexlab{b}})\citenamefont{{Hadzhiyska}, {Ferraro}, {Ried Guachalla}, {Schaan}, {Aguilar}, {Battaglia}, {Bond}, {Brooks}, {Calabrese}, {Choi} et~al.}}]{2407.07152}
\bibinfo{author}{\bibfnamefont{B.}~\bibnamefont{{Hadzhiyska}}}, \bibinfo{author}{\bibfnamefont{S.}~\bibnamefont{{Ferraro}}}, \bibinfo{author}{\bibfnamefont{B.}~\bibnamefont{{Ried Guachalla}}}, \bibinfo{author}{\bibfnamefont{E.}~\bibnamefont{{Schaan}}}, \bibinfo{author}{\bibfnamefont{J.}~\bibnamefont{{Aguilar}}}, \bibinfo{author}{\bibfnamefont{N.}~\bibnamefont{{Battaglia}}}, \bibinfo{author}{\bibfnamefont{J.~R.} \bibnamefont{{Bond}}}, \bibinfo{author}{\bibfnamefont{D.}~\bibnamefont{{Brooks}}}, \bibinfo{author}{\bibfnamefont{E.}~\bibnamefont{{Calabrese}}}, \bibinfo{author}{\bibfnamefont{S.~K.} \bibnamefont{{Choi}}}, \bibnamefont{et~al.}, \bibinfo{journal}{arXiv e-prints} \bibinfo{eid}{arXiv:2407.07152} (\bibinfo{year}{2024}{\natexlab{b}}), \eprint{2407.07152}.

\bibitem[{\citenamefont{{Hadzhiyska} et~al.}(2025)\citenamefont{{Hadzhiyska}, {Ferraro}, and {Zhou}}}]{2412.03631}
\bibinfo{author}{\bibfnamefont{B.}~\bibnamefont{{Hadzhiyska}}}, \bibinfo{author}{\bibfnamefont{S.}~\bibnamefont{{Ferraro}}}, \bibnamefont{and} \bibinfo{author}{\bibfnamefont{R.}~\bibnamefont{{Zhou}}}, \bibinfo{journal}{\prd} \textbf{\bibinfo{volume}{111}}, \bibinfo{eid}{023534} (\bibinfo{year}{2025}), \eprint{2412.03631}.

\bibitem[{\citenamefont{{White} et~al.}(2022)\citenamefont{{White}, {Zhou}, {DeRose}, {Ferraro}, {Chen}, {Kokron}, {Bailey}, {Brooks}, {Garc{\'\i}a-Bellido}, {Guy} et~al.}}]{White_2022}
\bibinfo{author}{\bibfnamefont{M.}~\bibnamefont{{White}}}, \bibinfo{author}{\bibfnamefont{R.}~\bibnamefont{{Zhou}}}, \bibinfo{author}{\bibfnamefont{J.}~\bibnamefont{{DeRose}}}, \bibinfo{author}{\bibfnamefont{S.}~\bibnamefont{{Ferraro}}}, \bibinfo{author}{\bibfnamefont{S.-F.} \bibnamefont{{Chen}}}, \bibinfo{author}{\bibfnamefont{N.}~\bibnamefont{{Kokron}}}, \bibinfo{author}{\bibfnamefont{S.}~\bibnamefont{{Bailey}}}, \bibinfo{author}{\bibfnamefont{D.}~\bibnamefont{{Brooks}}}, \bibinfo{author}{\bibfnamefont{J.}~\bibnamefont{{Garc{\'\i}a-Bellido}}}, \bibinfo{author}{\bibfnamefont{J.}~\bibnamefont{{Guy}}}, \bibnamefont{et~al.}, \bibinfo{journal}{\jcap} \textbf{\bibinfo{volume}{2022}}, \bibinfo{eid}{007} (\bibinfo{year}{2022}), \eprint{2111.09898}.

\bibitem[{\citenamefont{{Sailer} et~al.}(2024)\citenamefont{{Sailer}, {Kim}, {Ferraro}, {Madhavacheril}, {White}, {Abril-Cabezas}, {Aguilar}, {Ahlen}, {Bond}, {Brooks} et~al.}}]{sailer2024}
\bibinfo{author}{\bibfnamefont{N.}~\bibnamefont{{Sailer}}}, \bibinfo{author}{\bibfnamefont{J.}~\bibnamefont{{Kim}}}, \bibinfo{author}{\bibfnamefont{S.}~\bibnamefont{{Ferraro}}}, \bibinfo{author}{\bibfnamefont{M.~S.} \bibnamefont{{Madhavacheril}}}, \bibinfo{author}{\bibfnamefont{M.}~\bibnamefont{{White}}}, \bibinfo{author}{\bibfnamefont{I.}~\bibnamefont{{Abril-Cabezas}}}, \bibinfo{author}{\bibfnamefont{J.~N.} \bibnamefont{{Aguilar}}}, \bibinfo{author}{\bibfnamefont{S.}~\bibnamefont{{Ahlen}}}, \bibinfo{author}{\bibfnamefont{J.~R.} \bibnamefont{{Bond}}}, \bibinfo{author}{\bibfnamefont{D.}~\bibnamefont{{Brooks}}}, \bibnamefont{et~al.}, \bibinfo{journal}{arXiv e-prints} \bibinfo{eid}{arXiv:2407.04607} (\bibinfo{year}{2024}), \eprint{2407.04607}.

\bibitem[{\citenamefont{{Kim} et~al.}(2024)\citenamefont{{Kim}, {Sailer}, {Madhavacheril}, {Ferraro}, {Abril-Cabezas}, {Aguilar}, {Ahlen}, {Richard Bond}, {Brooks}, {Burtin} et~al.}}]{kim_2024}
\bibinfo{author}{\bibfnamefont{J.}~\bibnamefont{{Kim}}}, \bibinfo{author}{\bibfnamefont{N.}~\bibnamefont{{Sailer}}}, \bibinfo{author}{\bibfnamefont{M.~S.} \bibnamefont{{Madhavacheril}}}, \bibinfo{author}{\bibfnamefont{S.}~\bibnamefont{{Ferraro}}}, \bibinfo{author}{\bibfnamefont{I.}~\bibnamefont{{Abril-Cabezas}}}, \bibinfo{author}{\bibfnamefont{J.~N.} \bibnamefont{{Aguilar}}}, \bibinfo{author}{\bibfnamefont{S.}~\bibnamefont{{Ahlen}}}, \bibinfo{author}{\bibfnamefont{J.}~\bibnamefont{{Richard Bond}}}, \bibinfo{author}{\bibfnamefont{D.}~\bibnamefont{{Brooks}}}, \bibinfo{author}{\bibfnamefont{E.}~\bibnamefont{{Burtin}}}, \bibnamefont{et~al.}, \bibinfo{journal}{\jcap} \textbf{\bibinfo{volume}{2024}}, \bibinfo{eid}{022} (\bibinfo{year}{2024}), \eprint{2407.04606}.

\bibitem[{\citenamefont{{Farren} et~al.}(2024)\citenamefont{{Farren}, {Krolewski}, {MacCrann}, {Ferraro}, {Abril-Cabezas}, {An}, {Atkins}, {Battaglia}, {Bond}, {Calabrese} et~al.}}]{Farren_2024}
\bibinfo{author}{\bibfnamefont{G.~S.} \bibnamefont{{Farren}}}, \bibinfo{author}{\bibfnamefont{A.}~\bibnamefont{{Krolewski}}}, \bibinfo{author}{\bibfnamefont{N.}~\bibnamefont{{MacCrann}}}, \bibinfo{author}{\bibfnamefont{S.}~\bibnamefont{{Ferraro}}}, \bibinfo{author}{\bibfnamefont{I.}~\bibnamefont{{Abril-Cabezas}}}, \bibinfo{author}{\bibfnamefont{R.}~\bibnamefont{{An}}}, \bibinfo{author}{\bibfnamefont{Z.}~\bibnamefont{{Atkins}}}, \bibinfo{author}{\bibfnamefont{N.}~\bibnamefont{{Battaglia}}}, \bibinfo{author}{\bibfnamefont{J.~R.} \bibnamefont{{Bond}}}, \bibinfo{author}{\bibfnamefont{E.}~\bibnamefont{{Calabrese}}}, \bibnamefont{et~al.}, \bibinfo{journal}{\apj} \textbf{\bibinfo{volume}{966}}, \bibinfo{eid}{157} (\bibinfo{year}{2024}), \eprint{2309.05659}.

\bibitem[{\citenamefont{{Karim} et~al.}(2024)\citenamefont{{Karim}, {Singh}, {Rezaie}, {Eisenstein}, {Hadzhiyska}, {Speagle}, {Aguilar}, {Ahlen}, {Brooks}, {Claybaugh} et~al.}}]{karim2024desi}
\bibinfo{author}{\bibfnamefont{T.}~\bibnamefont{{Karim}}}, \bibinfo{author}{\bibfnamefont{S.}~\bibnamefont{{Singh}}}, \bibinfo{author}{\bibfnamefont{M.}~\bibnamefont{{Rezaie}}}, \bibinfo{author}{\bibfnamefont{D.}~\bibnamefont{{Eisenstein}}}, \bibinfo{author}{\bibfnamefont{B.}~\bibnamefont{{Hadzhiyska}}}, \bibinfo{author}{\bibfnamefont{J.~S.} \bibnamefont{{Speagle}}}, \bibinfo{author}{\bibfnamefont{J.~N.} \bibnamefont{{Aguilar}}}, \bibinfo{author}{\bibfnamefont{S.}~\bibnamefont{{Ahlen}}}, \bibinfo{author}{\bibfnamefont{D.}~\bibnamefont{{Brooks}}}, \bibinfo{author}{\bibfnamefont{T.}~\bibnamefont{{Claybaugh}}}, \bibnamefont{et~al.}, \bibinfo{journal}{arXiv e-prints} \bibinfo{eid}{arXiv:2408.15909} (\bibinfo{year}{2024}), \eprint{2408.15909}.

\bibitem[{\citenamefont{{Spergel} et~al.}(2015)\citenamefont{{Spergel}, {Gehrels}, {Baltay}, {Bennett}, {Breckinridge}, {Donahue}, {Dressler}, {Gaudi}, {Greene}, {Guyon} et~al.}}]{10.48550/arXiv.1503.03757}
\bibinfo{author}{\bibfnamefont{D.}~\bibnamefont{{Spergel}}}, \bibinfo{author}{\bibfnamefont{N.}~\bibnamefont{{Gehrels}}}, \bibinfo{author}{\bibfnamefont{C.}~\bibnamefont{{Baltay}}}, \bibinfo{author}{\bibfnamefont{D.}~\bibnamefont{{Bennett}}}, \bibinfo{author}{\bibfnamefont{J.}~\bibnamefont{{Breckinridge}}}, \bibinfo{author}{\bibfnamefont{M.}~\bibnamefont{{Donahue}}}, \bibinfo{author}{\bibfnamefont{A.}~\bibnamefont{{Dressler}}}, \bibinfo{author}{\bibfnamefont{B.~S.} \bibnamefont{{Gaudi}}}, \bibinfo{author}{\bibfnamefont{T.}~\bibnamefont{{Greene}}}, \bibinfo{author}{\bibfnamefont{O.}~\bibnamefont{{Guyon}}}, \bibnamefont{et~al.}, \bibinfo{journal}{arXiv e-prints} \bibinfo{eid}{arXiv:1503.03757} (\bibinfo{year}{2015}), \eprint{1503.03757}.

\bibitem[{\citenamefont{{The LSST Dark Energy Science Collaboration} et~al.}(2018)\citenamefont{{The LSST Dark Energy Science Collaboration}, {Mandelbaum}, {Eifler}, {Hlo{\v{z}}ek}, {Collett}, {Gawiser}, {Scolnic}, {Alonso}, {Awan}, {Biswas} et~al.}}]{1809.01669}
\bibinfo{author}{\bibnamefont{{The LSST Dark Energy Science Collaboration}}}, \bibinfo{author}{\bibfnamefont{R.}~\bibnamefont{{Mandelbaum}}}, \bibinfo{author}{\bibfnamefont{T.}~\bibnamefont{{Eifler}}}, \bibinfo{author}{\bibfnamefont{R.}~\bibnamefont{{Hlo{\v{z}}ek}}}, \bibinfo{author}{\bibfnamefont{T.}~\bibnamefont{{Collett}}}, \bibinfo{author}{\bibfnamefont{E.}~\bibnamefont{{Gawiser}}}, \bibinfo{author}{\bibfnamefont{D.}~\bibnamefont{{Scolnic}}}, \bibinfo{author}{\bibfnamefont{D.}~\bibnamefont{{Alonso}}}, \bibinfo{author}{\bibfnamefont{H.}~\bibnamefont{{Awan}}}, \bibinfo{author}{\bibfnamefont{R.}~\bibnamefont{{Biswas}}}, \bibnamefont{et~al.}, \bibinfo{journal}{arXiv e-prints} \bibinfo{eid}{arXiv:1809.01669} (\bibinfo{year}{2018}), \eprint{1809.01669}.

\bibitem[{\citenamefont{{Euclid Collaboration} et~al.}(2024)\citenamefont{{Euclid Collaboration}, {Mellier}, {Abdurro'uf}, {Acevedo Barroso}, {Ach{\'u}carro}, {Adamek}, {Adam}, {Addison}, {Aghanim}, {Aguena} et~al.}}]{2405.13491}
\bibinfo{author}{\bibnamefont{{Euclid Collaboration}}}, \bibinfo{author}{\bibfnamefont{Y.}~\bibnamefont{{Mellier}}}, \bibinfo{author}{\bibnamefont{{Abdurro'uf}}}, \bibinfo{author}{\bibfnamefont{J.~A.} \bibnamefont{{Acevedo Barroso}}}, \bibinfo{author}{\bibfnamefont{A.}~\bibnamefont{{Ach{\'u}carro}}}, \bibinfo{author}{\bibfnamefont{J.}~\bibnamefont{{Adamek}}}, \bibinfo{author}{\bibfnamefont{R.}~\bibnamefont{{Adam}}}, \bibinfo{author}{\bibfnamefont{G.~E.} \bibnamefont{{Addison}}}, \bibinfo{author}{\bibfnamefont{N.}~\bibnamefont{{Aghanim}}}, \bibinfo{author}{\bibfnamefont{M.}~\bibnamefont{{Aguena}}}, \bibnamefont{et~al.}, \bibinfo{journal}{arXiv e-prints} \bibinfo{eid}{arXiv:2405.13491} (\bibinfo{year}{2024}), \eprint{2405.13491}.

\bibitem[{\citenamefont{{Abazajian} et~al.}(2016)\citenamefont{{Abazajian}, {Adshead}, {Ahmed}, {Allen}, {Alonso}, {Arnold}, {Baccigalupi}, {Bartlett}, {Battaglia}, {Benson} et~al.}}]{1610.02743}
\bibinfo{author}{\bibfnamefont{K.~N.} \bibnamefont{{Abazajian}}}, \bibinfo{author}{\bibfnamefont{P.}~\bibnamefont{{Adshead}}}, \bibinfo{author}{\bibfnamefont{Z.}~\bibnamefont{{Ahmed}}}, \bibinfo{author}{\bibfnamefont{S.~W.} \bibnamefont{{Allen}}}, \bibinfo{author}{\bibfnamefont{D.}~\bibnamefont{{Alonso}}}, \bibinfo{author}{\bibfnamefont{K.~S.} \bibnamefont{{Arnold}}}, \bibinfo{author}{\bibfnamefont{C.}~\bibnamefont{{Baccigalupi}}}, \bibinfo{author}{\bibfnamefont{J.~G.} \bibnamefont{{Bartlett}}}, \bibinfo{author}{\bibfnamefont{N.}~\bibnamefont{{Battaglia}}}, \bibinfo{author}{\bibfnamefont{B.~A.} \bibnamefont{{Benson}}}, \bibnamefont{et~al.}, \bibinfo{journal}{arXiv e-prints} \bibinfo{eid}{arXiv:1610.02743} (\bibinfo{year}{2016}), \eprint{1610.02743}.

\bibitem[{\citenamefont{{Ade} et~al.}(2019)\citenamefont{{Ade}, {Aguirre}, {Ahmed}, {Aiola}, {Ali}, {Alonso}, {Alvarez}, {Arnold}, {Ashton}, {Austermann} et~al.}}]{1808.07445}
\bibinfo{author}{\bibfnamefont{P.}~\bibnamefont{{Ade}}}, \bibinfo{author}{\bibfnamefont{J.}~\bibnamefont{{Aguirre}}}, \bibinfo{author}{\bibfnamefont{Z.}~\bibnamefont{{Ahmed}}}, \bibinfo{author}{\bibfnamefont{S.}~\bibnamefont{{Aiola}}}, \bibinfo{author}{\bibfnamefont{A.}~\bibnamefont{{Ali}}}, \bibinfo{author}{\bibfnamefont{D.}~\bibnamefont{{Alonso}}}, \bibinfo{author}{\bibfnamefont{M.~A.} \bibnamefont{{Alvarez}}}, \bibinfo{author}{\bibfnamefont{K.}~\bibnamefont{{Arnold}}}, \bibinfo{author}{\bibfnamefont{P.}~\bibnamefont{{Ashton}}}, \bibinfo{author}{\bibfnamefont{J.}~\bibnamefont{{Austermann}}}, \bibnamefont{et~al.}, \bibinfo{journal}{\jcap} \textbf{\bibinfo{volume}{2019}}, \bibinfo{eid}{056} (\bibinfo{year}{2019}), \eprint{1808.07445}.

\bibitem[{\citenamefont{{MacInnis} and {Sehgal}}(2024)}]{2024arXiv240512220M}
\bibinfo{author}{\bibfnamefont{A.}~\bibnamefont{{MacInnis}}} \bibnamefont{and} \bibinfo{author}{\bibfnamefont{N.}~\bibnamefont{{Sehgal}}}, \bibinfo{journal}{arXiv e-prints} \bibinfo{eid}{arXiv:2405.12220} (\bibinfo{year}{2024}), \eprint{2405.12220}.

\bibitem[{\citenamefont{{Hunter}}(2007)}]{Hunter:2007}
\bibinfo{author}{\bibfnamefont{J.~D.} \bibnamefont{{Hunter}}}, \bibinfo{journal}{Computing in Science and Engineering} \textbf{\bibinfo{volume}{9}}, \bibinfo{pages}{90} (\bibinfo{year}{2007}).

\bibitem[{\citenamefont{{Lewis}}(2019)}]{1910.13970}
\bibinfo{author}{\bibfnamefont{A.}~\bibnamefont{{Lewis}}}, \bibinfo{journal}{arXiv e-prints} \bibinfo{eid}{arXiv:1910.13970} (\bibinfo{year}{2019}), \eprint{1910.13970}.

\bibitem[{\citenamefont{{Torrado} and {Lewis}}(2021)}]{2005.05290}
\bibinfo{author}{\bibfnamefont{J.}~\bibnamefont{{Torrado}}} \bibnamefont{and} \bibinfo{author}{\bibfnamefont{A.}~\bibnamefont{{Lewis}}}, \bibinfo{journal}{\jcap} \textbf{\bibinfo{volume}{2021}}, \bibinfo{eid}{057} (\bibinfo{year}{2021}), \eprint{2005.05290}.

\end{thebibliography}

\appendix
\section{Parameter correlations}
\label{section:app}

\begin{figure*}
    \centering
    \includegraphics[width=0.8 \textwidth]{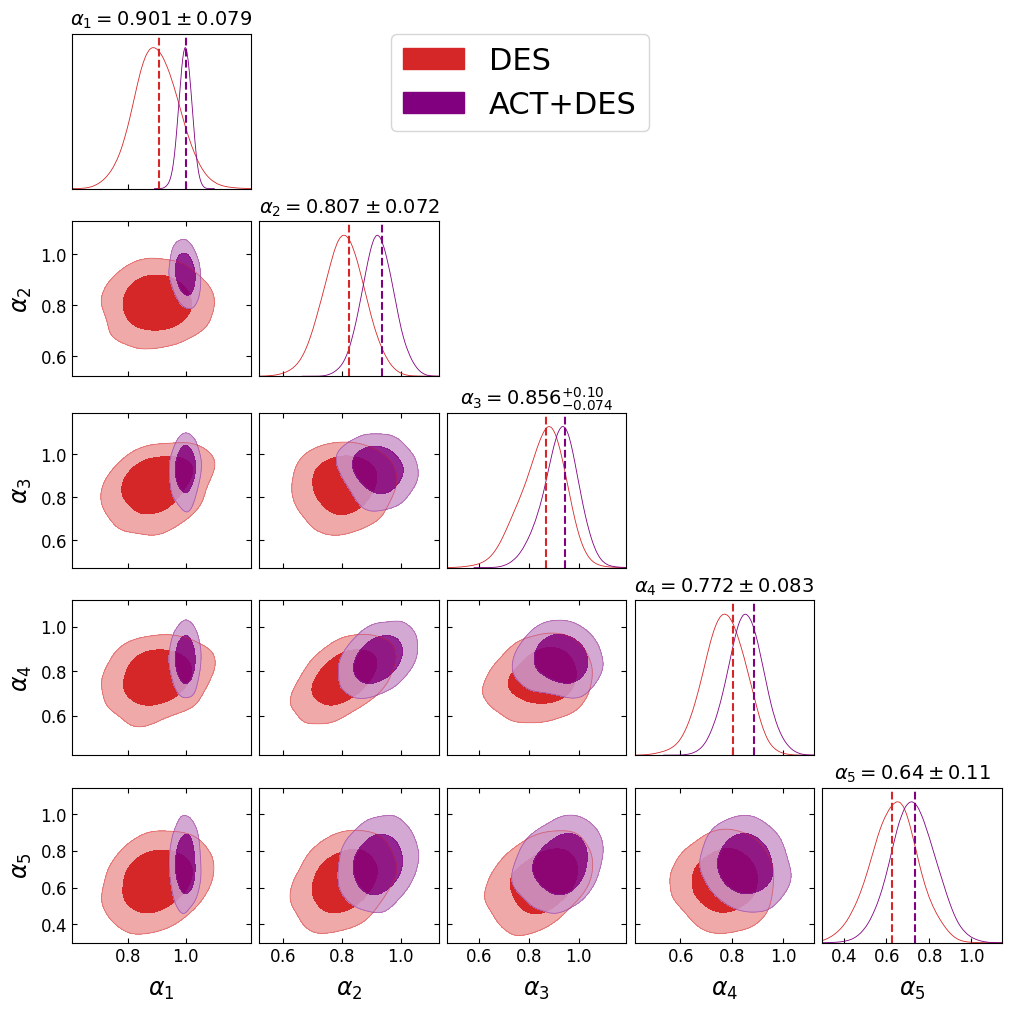}
    \caption{Marginalized 2d posteriors of the $\alpha_i$ parameters for the DES-only run and the ACT+DES run. Dashed vertical lines are the best-fit of the parameters.}
    \label{fig:alphas_triangle}
\end{figure*}

\begin{figure*}
    \centering
    \includegraphics[width=0.6 \textwidth]{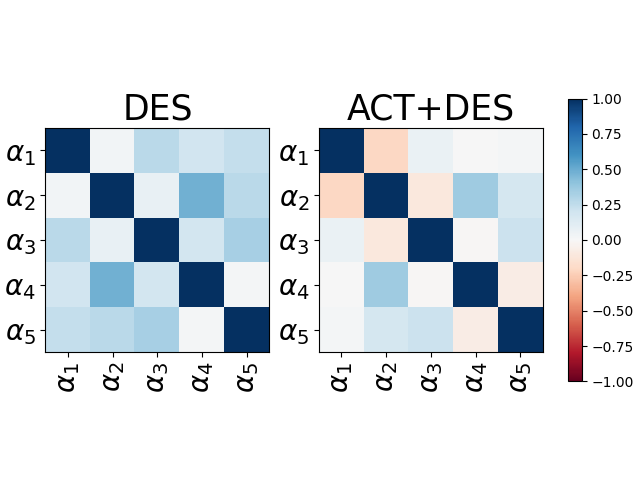}
    \vspace*{-17mm}
    \caption{Correlation matrices for the $\alpha_i$ parameters for the DES-only run (left) and the ACT+DES run (right).}
    \label{fig:alphas_corr}
\end{figure*}

Fig.~\ref{fig:alphas_triangle} shows the $\alpha_{i}$ parameters recovered from the DES-Y3 cosmic shear run and the joint run with ACT DR6 CMB lensing data, while Fig.~\ref{fig:alphas_corr} shows the corresponding correlation matrices. The best-fit values are generally in agreement with the mean values, and the correlation between different $\alpha_{i}$ parameters is generally low.

\end{document}